\documentclass[a4paper,12pt]{article}
\pdfoutput=1

\usepackage{jheppub}
\usepackage[utf8]{inputenc}
\usepackage{microtype}
\usepackage{booktabs}
\usepackage[svgnames]{xcolor}
\graphicspath{{figures/}}
\hypersetup{linkcolor=DarkBlue,citecolor=DarkBlue,urlcolor=black,
  pdftitle={Thermal deep inelastic scattering, universality and complex geodesics},
  pdfauthor={Andrei Parnachev}}
\makeatletter
\def\@fpheader{\vspace{1mm}}
\makeatother
\def\baselinestretch{1.14}

\let\paperTableOfContents\tableofcontents
\renewcommand{\tableofcontents}{%
  \begingroup
  \def\baselinestretch{0.90}\footnotesize
  \paperTableOfContents
  \endgroup
}

\newcommand{\ut}{\tilde u}

\renewcommand{\Im}{\operatorname{Im}}
\renewcommand{\Re}{\operatorname{Re}}
\newcommand{\dd}{\mathrm{d}}

\title{Thermal deep inelastic scattering,\\
universality and complex geodesics}
\author[a,b]{Andrei Parnachev}
\affiliation[a]{School of Mathematics and Hamilton Mathematics Institute,
  Trinity College Dublin, Dublin 2, Ireland}
\affiliation[b]{School of Mathematical Sciences, Queen Mary University of London,
  London, E1 4NS, United Kingdom}
\emailAdd{parnachev@maths.tcd.ie}

\abstract{In  deep inelastic scattering (DIS), a lepton scatters off a target hadron 
whose structure  is probed by an exchanged photon with large spacelike virtuality.
We consider two-point functions of currents   in strongly interacting quantum field theories at finite temperature 
in this kinematic regime and use a holographic description to compute them.
At large virtuality, the crossover from vacuum-dominated to thermally dominated physics corresponds to a transition from exponentially suppressed absorption to power-law behavior of the response functions. 
This  regime is related to a particular near-lightcone limit of thermal CFT correlators. 
We show that the response functions exhibit a large degree of universality: the dependence on the three-point couplings of currents to the stress tensor enters through a channel-dependent coefficient multiplying the kinematic parameter that controls the relative size of vacuum and thermal effects. 
The same coefficient controls  conformal collider bounds;
when either bound is saturated, the corresponding response function takes its vacuum form in the near-lightcone limit. Furthermore, the absorption exponent in the low-temperature regime is related to a complex geodesic propagating in the near-lightcone geometry.}

\begin{document}
\maketitle
\clearpage

\section{Introduction}
\label{sec:intro}

In deep inelastic scattering (DIS), a target is probed by a virtual photon with large spacelike virtuality.
We study an analogous kinematics for scattering off a plasma in thermal equilibrium, which we refer to as thermal DIS. 
The quantity of interest is the imaginary part of the retarded current two-point function: it measures absorption by the medium. 
There are two independent response functions, for polarizations transverse and longitudinal to the spatial momentum of the probe. 
The central question of this paper is how these two responses depend on the interaction between the current and the thermal state.

Ref.~\cite{Hatta:2007cs} computed this observable for an R-current of strongly coupled $\mathcal N=4$ supersymmetric Yang--Mills theory\footnote{
Holographic DIS was initiated in Ref.~\cite{Polchinski:2002jw}.
For a selection of works on holographic DIS, see Refs.~\cite{
  Brower:2006ea,
  Hatta:2007he,
  BallonBayona:2007qr,
  BallonBayona:2007rs,
  Cornalba:2008sp,
  Pire:2008zf,
  Albacete:2008ze,
  BallonBayona:2008zi,
  Levin:2009vj,
  Mueller:2008bt,
  Gao:2009ze,
  Hatta:2009ra,
  Avsar:2009xf,
  Cornalba:2009ax,
  Bayona:2009qe,
  Iancu:2009py,
  Hassanain:2009xw,
  Cornalba:2010vk,
  Kovchegov:2010uk,
  Brower:2010wf,
  Koile:2011aa,
  Stoffers:2012zw,
  Amorim:2018yod}.
  Thermal current two-point functions have also been studied in a variety of holographic models
  \cite{
  Policastro:2002se,
  Nunez:2003eq,
  Kovtun:2005ev,
  Teaney:2006nc,
  Caron-Huot:2006pee,
  Myers:2007we,
  Amado:2007pv,
  Amado:2007yr,
  Hatta:2008tx,
  Amado:2008ji,
  Arnold:2010ir,
  Hassanain:2012uj,
  Waeber:2018bea,
  He:2023wcs,
  Jia:2024zes}.
}.
The R-current is associated with an internal global symmetry. In the dual gravitational description, the R-current is represented by a gauge field in a black-brane geometry. Absorption by the plasma corresponds to incoming boundary conditions at the horizon.
The DIS regime of Ref.~\cite{Hatta:2007cs} involves large spacelike momentum transfer (the virtuality of the photon coupled to the $U(1)_R$ current).
In this situation, the bulk equation of motion, which is essentially a Schr\"odinger-type equation, features a potential with two competing terms.
The spacelike virtuality produces a potential barrier, whereas increasing the temperature makes the barrier easier to cross.
At fixed probe energy and virtuality, absorption at sufficiently low temperature is exponentially suppressed by tunnelling through the barrier. As the temperature increases, this suppression weakens and absorption becomes strong.
This competition was studied in Ref.~\cite{Mueller:2008bt}.

In a separate development, Ref.~\cite{Fitzpatrick:2019zqz} introduced an efficient way of computing thermal coefficients of
multi-stress tensors contributing to holographic two-point functions.
It was pointed out that the thermal coefficients of leading-twist multi-stress tensors, i.e. spin-$2n$ $n$-stress tensors, are universal -- they do not
change under the addition of higher-derivative terms to the bulk gravitational Lagrangian.
Such universality governs the near-lightcone behavior of thermal correlators and is a property that relies solely on the spectrum of operators in the theory and can be shown using the bootstrap
\cite{Karlsson:2019qfi,
  Kulaxizi:2019tkd,
  Karlsson:2019dbd,
  Li:2019zba}.\footnote{Near-lightcone correlators also exhibit an interesting mathematical structure, resembling vacuum blocks in two-dimensional CFTs with higher spin charges.
  For more work on near-lightcone correlators, see Refs.~
\cite{
  Li:2019tpf,
  Huang:2019fog,
  Fitzpatrick:2019efk,
  Karlsson:2019txu,
  Huang:2020ycs,
  Karlsson:2020ghx,
  Li:2020dqm,
  Parnachev:2020fna,
  Fitzpatrick:2020yjb,
  Parnachev:2020zbr,
  Belin:2020lsr,
  Besken:2020snx,
  Karlsson:2021duj,
  Huang:2021hye,
  Karlsson:2021mgg,
  Huang:2022vcs,
  Karlsson:2022osn,
  Dodelson:2022yvn,
  Huang:2022vet,
  Huang:2023ikg,
  Huang:2024wbq,
  Haehl:2025ehf,
  Huang:2026jbv}.
  }

As we show below, the DIS regime discussed in Refs.~\cite{Hatta:2007cs,Mueller:2008bt} is, in fact,  equivalent to the near-lightcone regime, where only the leading-twist multi-stress tensors and the operators degenerate with them
survive.
Hence it makes sense to ask questions about near-lightcone correlators of conserved currents.
One natural question is their universality.
A similar question was studied for the thermal holographic correlators of stress tensors in Ref.~\cite{Esper:2023jeq}, where it was pointed out that
the addition of higher-derivative gravitational terms cannot leave the near-lightcone correlator completely universal, because it modifies the cubic
stress-tensor coupling.
However, as shown in Ref.~\cite{Esper:2023jeq}, such couplings enter the near-lightcone correlator in a very simple way -- they simply rescale
the arguments of the response functions (there are three independent response functions in the stress-tensor case, as there are three independent polarizations).

In this paper we show that a similar scenario occurs for current correlators.
We add a coupling to the Weyl tensor while keeping the bulk geometry fixed~\cite{Hofman:2008ar,Ritz:2008kh,Hofman:2009ug}.
This is a way to change the current--current--stress-tensor three-point coupling in the boundary theory.
Our result is that, in the near-lightcone limit, the additional interaction only changes the response functions via a rescaling of the argument,
similarly to what happens in the stress-tensor case.

There is a separate response function and a separate rescaling for each polarization.
The two rescaling factors are the same combinations that govern the angular distribution of the energy flux in a state created by the current~\cite{Hofman:2008ar}.
Positivity of the energy flux gives the conformal collider bounds \cite{Hofman:2008ar}.
When either of the bounds is saturated, the corresponding near-lightcone current response function takes its vacuum form.
Its spacelike DIS absorption therefore vanishes.
This extends the ``freedom near lightcone'' phenomenon of Ref.~\cite{Huang:2022vet}, established there for stress-tensor correlators.
While the bootstrap derivation has only been carried out for scalars, 
the universality for stress tensors and currents can also be expected to follow from a bootstrap-type argument, once the spectrum is fixed.

This behavior has a simple geometric meaning.
In the near-lightcone limit the black brane reduces to a plane wave on anti-de Sitter space, the Kaigorodov spacetime, which is dual to the plasma in its infinite-momentum frame. 
The plasma enters the action of the gauge field  through the single component of the metric that carries its null energy density. 
The Weyl interaction can  only rescale the two coefficients of that  term. 

One may wonder whether QCD-like theories would correspond to some values of the current--current--stress-tensor coupling.
It would be interesting to perform the corresponding lattice calculations, particularly for larger values of the gauge group rank.

The absorptive exponent in the response function is very reminiscent of the exponentially suppressed terms
in the response function that reflect the bouncing singularity, a complex-time singularity in the correlator associated with the
null geodesic which bounces off the black hole singularity~\cite{Fidkowski:2003nf,Festuccia:2005pi,Ceplak:2024bja}.
The Borel resummation of the multi-stress tensor OPE in Ref.~\cite{Mueller:2008bt} and the subsequent WKB matching are reminiscent of
what happens in Ref.~\cite{Afkhami-Jeddi:2025wra}.\footnote{
Other recent papers devoted to bouncing singularities include
\cite{Dodelson:2023vrw,
  Horowitz:2023ury,
  Dodelson:2023nnr,
  Parisini:2023nbd,
  Faruk:2025bed,
  Buric:2025anb,
  Buric:2025fye,
  Barrat:2025twb,
  Dodelson:2025jff,
  Ceplak:2025dds,
  Chakravarty:2025ncy,
  Jia:2025jbi,
  AliAhmad:2026wem,
  Jia:2026pmv,
  Araya:2026shz,
  Giombi:2026kdz,
  Arnaudo:2026der,
  Grozdanov:2026cut,
  Jia:2026ryl,
  Arnaudo:2026tcy,
  Grozdanov:2026ktq,
  Buric:2026qsp,
  Buric:2026cmn,
  Arnaudo:2026axe,
  Barrat:2026jfg,
  Grozdanov:2026lnc,
  Hartnoll:2026vhu,
  Xiao:2026pir,
  Coviello:2026tol,
  Ceplak:2026ard,
  Coviello:2026hdc,
  Komargodski:2026oob,
  Batra:2026gbq,
  Buric:2026sit,
  Batra:2026otu,
  Chakravarty:2026fdm,
  Chen:2026efa,
  Almheiri:2026cpv,
  Oh:2026dky,
  Dodelson:2026hjm,
  Grassi:2026ekz,
  Barrat:2026ozj,
  Liu:2026luf,
  Arnaudo:2026gtd,
  Singhi:2026xbc,
  Oh:2026qcm,
  Jia:2026qvp}.
}
Indeed, one can show the direct relation by analyzing the near-lightcone equation in the low-temperature regime using the WKB approach.
In the near-lightcone limit the equations of the current become the equations of motion of the gauge field in the near-lightcone geometry of Ref.~\cite{Parnachev:2020fna}, and the leading WKB momentum obeys the null-geodesic equation in that geometry.
We compute the complex boundary displacements of the trajectory and obtain a purely imaginary eikonal phase.
Its exponential matches the current's WKB absorption factor, including its numerical coefficient and momentum dependence.
Note that these same complex endpoint separations give the scalar stress-sector singularity location studied in Ref.~\cite{Araya:2026shz}.
This relates the DIS exponent to that position-space result of Ref.~\cite{Araya:2026shz}.

The rest of the paper is organized as follows.
In Section~\ref{sec:dictionary} we review thermal DIS and establish its relation to the near-lightcone limit.
We explain how the vacuum barrier and the thermal effects compete, show that this happens precisely in the near-lightcone regime, and map the corresponding DIS and lightcone parameters. We also review the calculation of the absorption exponent in the low-temperature regime.
In Section~\ref{sec:currents} we introduce the Weyl interaction,
show that the response functions are related to those in the absence of this interaction by a simple rescaling of the argument and
 compute the two functions numerically across the crossover from strong to weak thermal absorption.
 We also highlight the Kaigorodov spacetime interpretation of the near-lightcone geometry.
 In Section~\ref{sec:geodesic} we match the absorption exponent to the (complex) eikonal phase computed from the complex null geodesic
  and relate it to the known position-space singular point.
  In Section~\ref{sec:discussion} we discuss our results and outline some questions.
  Appendix~\ref{app:reduced} derives the longitudinal equation,
  Appendix~\ref{app:numerics} describes the numerical calculation at vanishing Weyl coupling,
  Appendix~\ref{app:matching} gives the WKB matching at the turning point and near the boundary,
  and Appendix~\ref{app:moments} relates the thermal expansion coefficients to positive moments of the structure functions.

\section{Thermal DIS and the near-lightcone limit}
\label{sec:dictionary}

\subsection{DIS kinematics and structure functions}
\label{sec:observable}

Thermal deep inelastic scattering (DIS) probes how a plasma absorbs a spacelike current excitation. Ref.~\cite{Hatta:2007cs} studied this question for an R-current in strongly coupled $\mathcal N=4$ supersymmetric Yang--Mills (SYM) theory, using the holographic description of DIS developed in Refs.~\cite{Polchinski:2002jw,Hatta:2007he}. We first review that calculation and its energy dependence. We then identify the DIS limit that gives the near-lightcone equations.

Let $T$ be the temperature. In the plasma rest frame, the probe transfers energy $\omega>0$ and spatial momentum $k>0$ along the $z$ direction. Its spacelike virtuality $Q^2$ and Bjorken variable $x$ are
\begin{equation}
Q^2=k^2-\omega^2>0,\qquad
x=\frac{Q^2}{2\omega T}.
\label{eq:xQ}
\end{equation}
The momentum scale $Q$ sets the spatial resolution, of order $1/Q$. We choose $Q\gg T$ to probe distances shorter than the thermal scale $1/T$. At a given $Q$ and $T$, varying $\omega$ scans the plasma response: increasing $\omega$ decreases $x$. 

Ref.~\cite{Hatta:2007cs} considers $x$ of order one or smaller, with particular emphasis on small $x$. In this range, the definition \eqref{eq:xQ} and the hard-probe condition $Q\gg T$ imply
\begin{equation}
\frac{\omega}{Q}=\frac{Q}{2xT}\gg1,
\qquad k\simeq\omega\gg Q\gg T.
\label{eq:DIS-hierarchy-physical}
\end{equation}
This is the initial DIS hierarchy. It leaves the energy $\omega$ free to vary over a wide range. Expressing energy, momentum and virtuality in units of $2\pi T$, we define
\begin{equation}
w=\frac{\omega}{2\pi T},\qquad
q=\frac{k}{2\pi T},\qquad
K^2=q^2-w^2=\frac{Q^2}{(2\pi T)^2}.
\label{eq:dimensionless-momenta}
\end{equation}
The hierarchy \eqref{eq:DIS-hierarchy-physical} is therefore
\begin{equation}
q^2\gg K^2\gg1.
\label{eq:DIS-hierarchy}
\end{equation}
We will determine which combination of $q$ and $K$ controls absorption, before choosing a further scaling limit.

The observable is the retarded two-point function of the conserved current $J_\mu$. We use the mostly-plus metric $\eta_{\mu\nu}=\operatorname{diag}(-1,1,1,1)$ and the plasma four-velocity $n^\mu=(1,0,0,0)$; boundary indices $\mu,\nu$ run from $0$ to $3$. Writing $q^\mu$ for the dimensionful four-momentum and $\beta=1/T$, our convention is
\begin{equation}
R_{\mu\nu}(q)=i\int d^4x\,e^{-iq\cdot x}\theta(x^0)
\langle[J_\mu(x),J_\nu(0)]\rangle_\beta,
\qquad q^\mu=(\omega,0,0,k).
\label{eq:Rmunu}
\end{equation}
Here $x^\mu$ is a boundary spacetime coordinate, $\theta$ is the step function, and $\langle\cdot\rangle_\beta$ is a thermal expectation value. The scalar $q$ in \eqref{eq:dimensionless-momenta} remains the dimensionless spatial momentum. The imaginary part of $R_{\mu\nu}$ measures absorption by the plasma.

There are two independent polarization channels. The currents along $x$ and $y$ form the transverse channel, with equal responses by rotational symmetry. The charge density and the current along $z$ form the longitudinal channel, with their responses related by current conservation. The correlator is specified by two scalar functions, $R_1$ and $R_2$:
\begin{equation}
R_{\mu\nu}=\left(\eta_{\mu\nu}-\frac{q_\mu q_\nu}{Q^2}\right)R_1
+\left(n_\mu-\frac{n\!\cdot\!q}{Q^2}q_\mu\right)
 \left(n_\nu-\frac{n\!\cdot\!q}{Q^2}q_\nu\right)R_2.
\label{eq:decomp}
\end{equation}
Both functions depend on $Q$, $x$ and $T$. The conventional structure functions $F_1$ and $F_2$ are defined by their imaginary parts. The combinations $F_T$ and $F_L$ separate transverse and longitudinal absorption in the DIS regime:
\begin{equation}
\begin{aligned}
F_1&=\frac{1}{2\pi}\,\mathrm{Im}\,R_1,
& F_2&=\frac{\omega}{2\pi T}\,\mathrm{Im}\,R_2,\\
F_T&=2xF_1,
& F_L&=F_2-2xF_1.
\end{aligned}
\label{eq:F12}
\end{equation}
The tensor multiplying $R_1$ in \eqref{eq:decomp} contributes to both polarizations. The longitudinal projection therefore involves both $F_1$ and $F_2$, giving $F_L=F_2-2xF_1$ in the DIS limit \eqref{eq:DIS-hierarchy-physical}. These thermal structure functions have mass dimension two.

\subsection{The two radial equations at strong coupling}
\label{sec:thermal-DIS}

For gauge group $SU(N)$, Ref.~\cite{Hatta:2007cs} takes both $N$ and the 't Hooft coupling $\lambda=g_{\rm YM}^2N$ to be large, where $g_{\rm YM}$ is the Yang--Mills coupling. The dual probe is a gauge field in a planar anti-de Sitter black-brane geometry. With $L$ the anti-de Sitter radius, the metric is
\begin{equation}
\begin{aligned}
ds^2&=\frac{(\pi TL)^2}{u}\big[-f(u)\,dt^2+d\mathbf x^2\big]
+\frac{L^2}{4u^2f(u)}\,du^2,\\
f(u)&=1-u^2.
\end{aligned}
\label{eq:metric}
\end{equation}
The dimensionless radial coordinate $u$ runs from the boundary at $u=0$ to the horizon at $u=1$. In terms of the usual radial coordinate $r$, it is $u=r_+^2/r^2$, where $r_+=\pi L^2T$ is the horizon radius. The boundary gauge field specifies the applied current source. We obtain the retarded correlator by imposing incoming boundary conditions at the horizon \cite{Son:2002sd,Policastro:2002se,Kovtun:2005ev}.

We use the standard current normalization of Ref.~\cite{Hatta:2007cs}. In terms of the five-dimensional gauge coupling $g^2$, which enters the bulk gauge field action through its overall factor $1/(4g^2)$, this convention is
\begin{equation}
\frac{L}{g^2}=\frac{N^2}{16\pi^2}.
\label{eq:current-normalization}
\end{equation}
The coupling $g^2$ is distinct from the Yang--Mills coupling $g_{\rm YM}^2$. It fixes the overall normalization of $R_{\mu\nu}$ and the structure functions.

We choose radial gauge and Fourier modes with the same $\omega$ and $k$ as the probe:
\begin{equation}
A_u=0,\qquad
A_\mu(t,\mathbf x,u)=e^{-i\omega t+ikz}A_\mu(u).
\label{eq:gauge-ansatz}
\end{equation}
The transverse field $A_\perp$ and longitudinal variable $a$ are
\begin{equation}
A_\perp(u)=A_x(u),\qquad a(u)=\frac{dA_t(u)}{du}.
\label{eq:current-masterfields}
\end{equation}
The field $A_y$ obeys the same equation as $A_x$. In the longitudinal channel, the equations of motion couple $A_t$ and $A_z$. With a prime denoting a $u$ derivative, their radial constraint is
\begin{equation}
wA_t'+qfA_z'=0.
\label{eq:gauge-constraint}
\end{equation}
Differentiating the temporal equation of motion and using \eqref{eq:gauge-constraint} eliminates $A_z'$ in favor of $a$. The two independent equations are then \cite{Hatta:2007cs}
\begin{align}
A_\perp''+\frac{f'}{f}A_\perp'
 +\frac{w^2-q^2f}{uf^2}A_\perp&=0,\nonumber\\
a''+\frac{(uf)'}{uf}a'
 +\frac{w^2-q^2f}{uf^2}a&=0.
\label{eq:current-full}
\end{align}
These equations are exact within the classical gauge-field description. Their different first-derivative terms distinguish the two polarizations.

\subsubsection{Exact potentials and the large-virtuality approximation}

We remove the first derivatives in \eqref{eq:current-full} by defining the fields $\Phi_T$ and $\Phi_L$. The resulting equations define the radial potentials $V_T$ and $V_L$:
\begin{equation}
\Phi_T=\sqrt f\,A_\perp,\qquad
\Phi_L=\sqrt{uf}\,a,\qquad
\frac{d^2\Phi_i}{du^2}=V_i\Phi_i,\quad i=T,L.
\label{eq:Schrodinger-fields}
\end{equation}
Using $w^2-q^2f=-K^2+q^2u^2$, the potentials are
\begin{equation}
\begin{aligned}
V_T(u)&=\frac1{f(u)^2}\left(\frac{K^2}{u}-q^2u-1\right),\\
V_L(u)&=\frac1{f(u)^2}\left(\frac{K^2}{u}-q^2u-\frac1{4u^2}
                                      -\frac32+\frac34u^2\right).
\end{aligned}
\label{eq:exact-potentials}
\end{equation}
For $K\gg1$, one retains
\begin{equation}
\begin{aligned}
V_T(u)&\simeq\frac{K^2}{uf(u)^2}-\frac{q^2u}{f(u)^2},\\
V_L(u)&\simeq\frac{K^2}{uf(u)^2}-\frac{q^2u}{f(u)^2}
                          -\frac1{4u^2f(u)^2}.
\end{aligned}
\label{eq:schr}
\end{equation}
The omitted terms are $-1/f^2$ in $V_T$ and $-3(2-u^2)/(4f^2)$ in $V_L$. Relative to the retained virtuality term $K^2/(uf^2)$, their magnitudes are $u/K^2$ and $3u(2-u^2)/(4K^2)$, respectively. They are small for $0<u<1$ when $K\gg1$. The high-energy condition $q\gg K$ has a separate role, explained below.

The term $K^2/(uf^2)$ opposes penetration into the bulk. The term $-q^2u/f^2$ enhances penetration as the energy grows. We call these the virtuality and energy-enhanced thermal contributions. The longitudinal term $-1/(4u^2f^2)$ comes from the field redefinition \eqref{eq:Schrodinger-fields}. The factors of $f$ also depend on temperature, so the first term in \eqref{eq:schr} is not by itself the complete vacuum potential.

\subsubsection{The vacuum limit}

To compare thermal and vacuum propagation, use the radial variable
\begin{equation}
v=K^2u=\frac{Q^2L^4}{4r^2}.
\label{eq:fixed-v}
\end{equation}
Unlike $u$, the coordinate $v$ is independent of temperature at fixed physical virtuality $Q$ and radial position $r$. It therefore lets us take the vacuum limit without moving the physical radial point. Define the potentials in this coordinate by
\begin{equation}
\widehat V_i=\frac{V_i}{K^4},\qquad
\frac{d^2\Phi_i}{dv^2}=\widehat V_i\Phi_i,\quad i=T,L.
\label{eq:scaled-potentials}
\end{equation}

Take $T\to0$ at fixed $Q,k,\omega,r$. Then $u\to0$, $K\to\infty$ and $v$ remains fixed. Applying this limit to the exact potentials \eqref{eq:exact-potentials} gives
\begin{equation}
\widehat V_T^{\,\rm vac}(v)=\frac1v,\qquad
\widehat V_L^{\,\rm vac}(v)=\frac1v-\frac1{4v^2}.
\label{eq:vacuum-potentials}
\end{equation}
In particular, the longitudinal inverse-square term survives at zero temperature. For the transverse field, the vacuum equation can be written in either coordinate:
\begin{equation}
\frac{d^2A_\perp}{du^2}=\frac{K^2}{u}A_\perp,
\qquad
\frac{d^2A_\perp}{dv^2}=\frac1v A_\perp.
\label{eq:vacuum-radial-scale}
\end{equation}
The virtuality $Q$ is still present through $v=K^2u$ in \eqref{eq:fixed-v}. In the variable $v$, all virtualities have the same normalized vacuum profile. The profile changes appreciably at $v\sim1$ and decays as $\exp(-2\sqrt v)$ for $v\gg1$, up to powers of $v$. The longitudinal profile has the same large-$v$ exponential. Thus the vacuum decay scale in the original coordinate is $u\sim K^{-2}\ll1$ when $Q\gg T$.

Both potentials in \eqref{eq:vacuum-potentials} have a positive $1/v$ tail extending to infinity. Regularity selects decaying solutions, which carry no flux into the interior. The negative longitudinal potential near $v=0$ does not change this conclusion: the local solutions are 
\begin{equation}
\Phi_L(v)\sim\sqrt v\big(C_1+C_2\log v\big),\qquad v\to0,
\label{eq:vacuum-longitudinal-boundary}
\end{equation}
where $C_1$ and $C_2$ are integration constants.

The absence of spacelike vacuum absorption also follows directly from boundary kinematics. A state created from the vacuum by $J_\mu$ would have energy $\omega$ and momentum $k$, but the spectrum condition requires $\omega^2\ge k^2$. Our probe has $k^2-\omega^2=Q^2>0$.
 With subtraction scale $\Lambda$, the result is \cite{Hatta:2007cs}
\begin{equation}
R_1^{(0)}=\frac{N^2Q^2}{32\pi^2}\ln\frac{Q^2}{\Lambda^2}.
\label{eq:vacuum-polarization}
\end{equation}

\subsubsection{Virtuality and thermal absorption}

At nonzero temperature, compare the two momentum-dependent terms in \eqref{eq:schr}. Their ratio is independent of $f$:
\begin{equation}
\frac{q^2u/f^2}{K^2/(uf^2)}=\frac{q^2u^2}{K^2}.
\label{eq:thermal-virtuality-ratio}
\end{equation}
They balance at $u\sim K/q$. The DIS condition $q\gg K$ in \eqref{eq:DIS-hierarchy} places this scale near the boundary. The vacuum decay scale $u\sim K^{-2}$ is also near the boundary. We can therefore compare the decay of the vacuum field with the growth of the thermal term using $f\simeq1$ in this region.

After absorbing a constant normalization in $\Phi_L$, the fields \eqref{eq:Schrodinger-fields} become $\Phi_T=A_\perp$ and $\Phi_L=\sqrt v\,a$ at leading order near the boundary. Introduce the dimensionless parameter used in Ref.~\cite{Mueller:2008bt},
\begin{equation}
\rho=\frac{q^2}{32K^6}
=\frac{k^2(2\pi T)^4}{32Q^6}.
\label{eq:thermal-rho}
\end{equation}
Equations \eqref{eq:schr} and \eqref{eq:scaled-potentials} then give
\begin{equation}
\begin{aligned}
\widehat V_T(v)&\simeq\frac1v-32\rho v,\\
\widehat V_L(v)&\simeq\frac1v-\frac1{4v^2}-32\rho v.
\end{aligned}
\label{eq:reduced-potentials-v}
\end{equation}
For fixed positive $v$ and fixed $\rho$, the terms omitted from \eqref{eq:reduced-potentials-v} are of order $K^{-4}$. This follows from $f=1-v^2/K^4$ and the exact potentials \eqref{eq:exact-potentials}.

At $v\sim1$, where the vacuum field begins to decay appreciably, the thermal contribution relative to the vacuum contribution is of order $\rho$. Thus $\rho$ controls the crossover between weak and strong thermal absorption.

The same criterion has a direct DIS interpretation. From \eqref{eq:xQ}, \eqref{eq:dimensionless-momenta} and \eqref{eq:thermal-rho},
\begin{equation}
\sqrt{\rho}\sim\frac q{K^3}\sim\frac{T}{xQ}.
\label{eq:thermal-control}
\end{equation}
Here numerical factors have been omitted and $k\simeq\omega$ from \eqref{eq:DIS-hierarchy-physical} has been used. The crossover between weak and strong thermal absorption occurs at $x\sim T/Q$.

\begin{figure}[t]
\centering
\includegraphics[width=\textwidth]{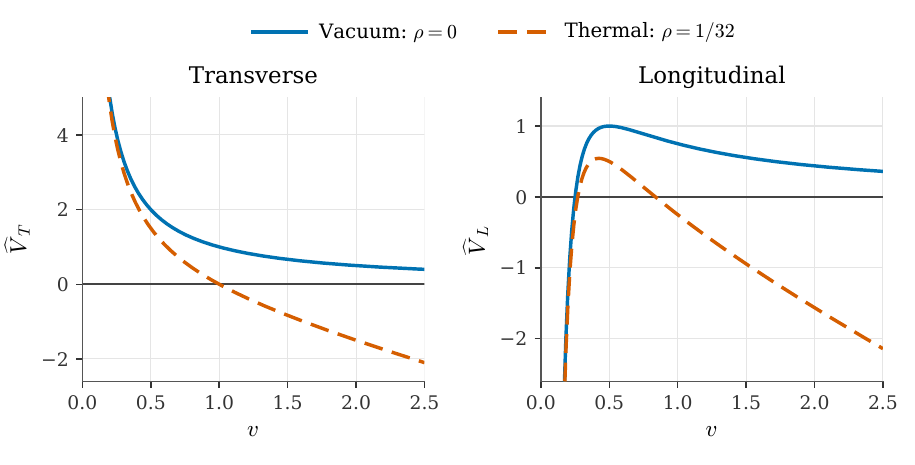}
\caption{The reduced potentials \eqref{eq:reduced-potentials-v} for $\rho=0$ and $\rho=1/32$. The panels use different vertical scales, and the boundary divergences extend beyond the plotted ranges. The thermal term $-32\rho v$ opens a propagation region at large $v$. The negative longitudinal region near $v=0$ instead has the nonoscillatory behavior \eqref{eq:vacuum-longitudinal-boundary}. The full geometry near the horizon is not shown.}
\label{fig:current-potentials}
\end{figure}

\subsection{Weak and strong thermal absorption}
\label{sec:fixed-rho-DIS}

We keep $\rho$ fixed as $K$ becomes large, so that the relative size of the thermal and vacuum contributions remains fixed.

The reduced equations inherit the incoming boundary condition at the horizon. In the large-$K$ limit, the horizon is at $v=\infty$, so this condition is imposed at large $v$ and selects propagation toward increasing $v$. Ref.~\cite{Hatta:2007cs} obtains this prescription by matching the small-$u$ solution to the solution deeper in the geometry.

\subsubsection{Weak thermal absorption and the OPE}

For $\rho\ll1$, or parametrically $x\gg T/Q$, the term $-32\rho v$ in \eqref{eq:reduced-potentials-v} is small in the region $v\sim1$. Iterating the radial equations around their vacuum solutions gives a multiple-scattering expansion in $\rho$. Its term of order $\rho^n$ describes $n$ graviton exchanges with the black-brane background, where $n$ is a positive integer. The first thermal corrections to the scalar correlators are \cite{Hatta:2007cs}
\begin{equation}
\begin{aligned}
R_1^{(1)}&=\frac{N^2\pi^2T^2}{40x^2},\\
R_2^{(1)}&=\frac{N^2\pi^2T^4}{6Q^2}.
\end{aligned}
\label{eq:lt}
\end{equation}
The superscript $(1)$ labels the first thermal correction. It comes from the stress tensor $T_{\mu\nu}$ in the operator product expansion (OPE) of two currents, evaluated in the thermal state. Higher graviton exchanges generate the corresponding multi-stress-tensor contributions. In this calculation, the leading near-lightcone contributions organize into powers of $\rho$.

The corrections \eqref{eq:lt}, and every higher finite order of the same spacelike expansion, are real. Hence no finite order gives a nonzero absorptive structure function \cite{Hatta:2007cs,Mueller:2008bt}. The full retarded solution nevertheless has exponentially small absorption, because the thermal term eventually dominates at large $v$ even when $\rho$ is small.

\subsubsection{Tunnelling and the absorption exponent}

Consider the transverse potential in \eqref{eq:reduced-potentials-v}. It is positive between the boundary and the turning point $v_*$, and negative beyond it, where
\begin{equation}
v_*=\frac1{\sqrt{32\rho}}=\frac{K^3}{q}.
\label{eq:transverse-turning-point}
\end{equation}
For $\rho\ll1$, the interval $0<v<v_*$ is a wide barrier. The transverse field $A_\perp$, sourced at the boundary, must tunnel through the barrier before propagating toward the horizon. Figure~\ref{fig:current-potentials} illustrates the thermal change in the potentials.

The WKB approximation determines the leading tunnelling exponential for $\rho\ll1$, after taking $K$ large at fixed $\rho$. At finite $K$, the turning point must obey $v_*\ll K^2$, equivalently $q\gg K$, to remain near the boundary. Let $D_T$ be the attenuation of transverse flux through the barrier. The WKB integral gives
\begin{equation}
\log D_T=-2\int_0^{v_*}dv\,\sqrt{\frac1v-32\rho v}
+o\big(\rho^{-1/4}\big).
\label{eq:current-WKB-integral}
\end{equation}
The holographic prescription relates $\Im R_1$ to the conserved transverse radial flux, divided by the squared source amplitude $|A_\perp(0)|^2$. Evaluated beyond the barrier, this flux is proportional to the squared magnitude of the transmitted $A_\perp$ amplitude. The factor of two in \eqref{eq:current-WKB-integral} therefore comes from squaring the WKB transmission amplitude. Appendix~\ref{app:matching} gives the relation between the flux and the correlator.

WKB fails near $v=0$ and $v=v_*$. Extending the integral to these endpoints gives the correct leading exponent: matching to the local boundary and turning-point solutions changes only subleading terms. Appendix~\ref{app:matching} explains this matching.

Define $s=v/v_*$, so the barrier occupies $0<s<1$. The integral in \eqref{eq:current-WKB-integral} is
\begin{equation}
\begin{aligned}
2\int_0^{v_*}dv\,\sqrt{\frac1v-\frac{v}{v_*^2}}
&=2\sqrt{v_*}\int_0^1ds\,\sqrt{\frac1s-s}\\[5pt]
&=A\sqrt{\frac{v_*}{2}}.
\end{aligned}
\label{eq:current-WKB-scaled}
\end{equation}
With $B$ the Euler beta function and $\Gamma$ the Euler gamma function, the coefficient $A$ is
\begin{equation}
\begin{aligned}
A&=\sqrt2\,B\!\left(\frac14,\frac32\right)\\[5pt]
&=\frac{2\Gamma(1/4)^2}{3\sqrt\pi}
=4.9442\ldots .
\end{aligned}
\label{eq:current-WKB-constant}
\end{equation}

For the longitudinal potential in \eqref{eq:reduced-potentials-v}, the term $-1/(4v^2)$ is essential near the boundary. The vacuum longitudinal solution is first matched to a WKB solution in the overlap region
\begin{equation}
1\ll v\ll\rho^{-2/5}.
\label{eq:longitudinal-WKB-overlap}
\end{equation}
The leading action comes from $v\sim v_*$, where $-1/(4v^2)$ is subleading. The two polarizations therefore have the same leading absorption exponent. The boundary matching in Ref.~\cite{Hatta:2007cs} gives
\begin{equation}
F_i\sim N^2Q^2x\,
\exp\!\left[-\frac{A}{\sqrt2}\left(\frac{K^3}{q}\right)^{1/2}\right],
\qquad i=T,L.
\label{eq:tunnel}
\end{equation}
Here $\sim$ specifies the leading exponential dependence. Both absorptive contributions are smaller than every finite power of $\rho$.

\subsubsection{Strong thermal absorption and the saturation scale}

For $q/K^3\gg1$, or parametrically $x\ll T/Q$, attenuation is no longer strong. The virtuality term can be neglected in the radial region that controls absorption, and Ref.~\cite{Hatta:2007cs} finds
\begin{equation}
\begin{aligned}
F_1&=\frac{3N^2T^2}{16\Gamma(1/3)^2}\left(\frac q3\right)^{2/3},\\
F_L&=\frac{N^2Q^2x}{96\pi^2}.
\end{aligned}
\label{eq:highE}
\end{equation}
The transverse contribution dominates, giving $F_2\simeq2xF_1$. The saturation momentum $Q_s(x)$ separates weak from strong thermal absorption and satisfies parametrically
\begin{equation}
Q_s(x)\sim\frac{T}{x}.
\label{eq:Qs}
\end{equation}
Thus $Q\gg Q_s(x)$ gives weak thermal absorption, whereas $Q\ll Q_s(x)$ gives strong thermal absorption. Keeping $\rho$ fixed keeps $Q/Q_s(x)$ parametrically fixed, by \eqref{eq:thermal-control}. Ref.~\cite{Hatta:2007cs} interprets \eqref{eq:Qs} in terms of plasma partons concentrated at $x\lesssim T/Q$, with occupation numbers of order one below $Q_s$; see also Refs.~\cite{Hatta:2008tx,Iancu:2008sp}.

\subsubsection{Large-order OPE coefficients and absorption}
\label{sec:large-order-OPE}

Ref.~\cite{Mueller:2008bt} studied the large-order behavior of the thermal expansion in $\rho$. Let $a_n$ be its $n$-th longitudinal boundary coefficient, denoted $a_n(0)$ in that reference; the corresponding series term is proportional to $a_n\rho^n$. From coefficients through $n=13$, the authors inferred
\begin{equation}
\begin{aligned}
\frac{a_{n+1}}{a_n}&\sim\frac{128}{A^4}\frac{\Gamma(4n+4)}{\Gamma(4n)},\\[5pt]
a_n&\sim\Gamma(4n)\left(\frac{128}{A^4}\right)^n.
\end{aligned}
\label{eq:large-order-series}
\end{equation}
The coefficient ratios give a numerical estimate $A\simeq4.95$ in our notation, agreeing with the WKB value in \eqref{eq:current-WKB-constant}. The exact expression for $A$ used in \eqref{eq:large-order-series} was motivated by that WKB result.

The factorial growth in \eqref{eq:large-order-series} makes the thermal expansion divergent. Its generalized Borel representation has a singularity on the positive integration axis. Passing above or below this singularity gives a difference with exponential dependence
\begin{equation}
\exp\!\left[-\frac{A}{(128\rho)^{1/4}}\right]
=\exp\!\left[-\frac{A}{\sqrt2}\left(\frac{K^3}{q}\right)^{1/2}\right].
\label{eq:borel}
\end{equation}
This is the absorption factor in \eqref{eq:tunnel}. Although every finite order is real, the large-order OPE coefficients encode the exponentially small absorption.

The same connection between OPE data and exponentially small contributions appears in the study of bouncing singularities \cite{Ceplak:2024bja,Afkhami-Jeddi:2025wra,Giombi:2026kdz}. There, large-order OPE coefficients determine complex-time singularities that produce exponentially suppressed terms in the frequency-space correlator; WKB relates these terms to null geodesics. In Section~\ref{sec:geodesic}, we identify a complex null geodesic that reproduces the DIS suppression exponent \eqref{eq:borel}.

\subsection{Near-lightcone variables}
\label{sec:nearlc}

Keeping $\rho$ fixed retains the competition between thermal and vacuum propagation as the resolution increases. We now express it in the variables used for near-lightcone thermal correlators in Ref.~\cite{Esper:2023jeq}. Define the dimensionless lightcone momenta $q^\pm=w\pm q$. The spacelike scaling limit is
\begin{equation}
\begin{gathered}
q^+\to\infty,\qquad q^-\to0^-,\\
\alpha=-q^+(q^-)^3>0\ \text{fixed},\qquad
\tilde u=\frac{u}{(q^-)^2}\ \text{fixed}.
\end{gathered}
\label{eq:limit}
\end{equation}
Here $\alpha$ is a momentum parameter and $\tilde u$ is a rescaled radial coordinate. Their relation to the DIS variables is exact before taking the limit:
\begin{equation}
\begin{gathered}
K^2=-q^+q^-,\qquad x=\frac{\pi K^2}{w},\\
\alpha=\frac{K^6}{(q^+)^2}
=\frac{Q^6}{(2\pi T)^4(\omega+k)^2}.
\end{gathered}
\label{eq:dict1}
\end{equation}
In the scaling limit \eqref{eq:limit}, $w/q\to1$. The identities \eqref{eq:dict1} and \eqref{eq:thermal-rho} then give
\begin{equation}
\begin{aligned}
\alpha&\simeq\frac14\left(\frac{K^3}{q}\right)^2
\simeq\frac1{16\pi^4}\left(\frac{xQ}{T}\right)^2,\\
\rho&=\frac1{128\alpha}.
\end{aligned}
\label{eq:dict2}
\end{equation}
The relations \eqref{eq:dict2} hold in the scaling limit, whereas \eqref{eq:dict1} holds at finite momenta. In physical variables, fixed $\alpha$ means
\begin{equation}
\frac QT\to\infty,\qquad x\to0,\qquad
\frac{xQ}{T}\to4\pi^2\sqrt\alpha\quad\text{fixed}.
\label{eq:dictlimit}
\end{equation}
By \eqref{eq:dict2}, this is precisely the fixed-$\rho$ limit motivated by DIS. The ratio $Q/T$ grows even though the momentum approaches the lightcone, because the energy grows faster than $Q$.

To obtain the limiting current equations directly, introduce a positive parameter $\epsilon$ and write
\begin{equation}
q^-=-\epsilon,\qquad q^+=\frac{\alpha}{\epsilon^3},\qquad
u=\epsilon^2\tilde u,\qquad \epsilon\to0,
\label{eq:epsilon-limit}
\end{equation}
with $\alpha$ and $\tilde u$ fixed. In the exact equations \eqref{eq:current-full}, each second derivative carries a factor $\epsilon^{-4}$. Multiplying by $\epsilon^4$, the momentum term becomes
\begin{equation}
\epsilon^4\frac{w^2-q^2f}{uf^2}
=-\frac{\alpha}{\tilde u}+\frac{\alpha^2\tilde u}{4}+O(\epsilon^4).
\label{eq:current-limit-potential}
\end{equation}
The transverse first-derivative term vanishes in this limit. The longitudinal first-derivative term remains, and the two equations reduce to
\begin{align}
\frac{d^2A_\perp}{d\tilde u^2}
&=\left(\frac{\alpha}{\tilde u}-\frac{\alpha^2\tilde u}{4}\right)A_\perp,\nonumber\\
\frac{d^2a}{d\tilde u^2}+\frac1{\tilde u}\frac{da}{d\tilde u}
&=\left(\frac{\alpha}{\tilde u}-\frac{\alpha^2\tilde u}{4}\right)a.
\label{eq:reduced}
\end{align}
Both equations retain the virtuality term proportional to $\alpha$ and the thermal term proportional to $\alpha^2$. Their different first derivatives imply distinct transverse and longitudinal response functions. The earlier radial coordinate $v$ is related to $\tilde u$ by
\begin{equation}
v=\alpha\tilde u.
\label{eq:v-lightcone}
\end{equation}
Using \eqref{eq:v-lightcone} and the field redefinitions \eqref{eq:Schrodinger-fields}, the equations \eqref{eq:reduced} reproduce the limiting potentials \eqref{eq:reduced-potentials-v}.

To define the response functions, first specify the sources. Let $C_T=A_\perp(0)$ be the transverse boundary source. In the longitudinal channel, the source is
\begin{equation}
A_L(0)=A_t(0)+\frac wq A_z(0).
\label{eq:longitudinal-boundary-source}
\end{equation}
The field $a=dA_t/du$ has a boundary logarithm. Its source coefficient $C_L$ is fixed by the boundary constraint \cite{Hatta:2007cs}:
\begin{equation}
C_L=\lim_{v\to0}v\frac{da}{dv}
=q\big[qA_t(0)+wA_z(0)\big]=q^2A_L(0).
\label{eq:longitudinal-source}
\end{equation}
With these source coefficients fixed, the finite boundary terms define the dimensionless response functions $f_T(\alpha)$ and $f_L(\alpha)$:
\begin{align}
A_\perp(v)&=C_T\big[1+v\log v+f_T(\alpha)v
+O(v^2\log v)\big],\nonumber\\
a(v)&=C_L\big[\log v-f_L(\alpha)+O(v\log v)\big].
\label{eq:current-boundary}
\end{align}
The incoming condition at large $v$, inherited from the retarded black-brane problem, determines $f_T(\alpha)$ and $f_L(\alpha)$. Their imaginary parts give the absorption; their real parts include the perturbative thermal corrections. The large-$\alpha$ expansion is the multiple-scattering expansion in powers of $\rho$, rewritten using \eqref{eq:dict2}.

With the current normalization \eqref{eq:current-normalization}, the boundary coefficients \eqref{eq:current-boundary} give the DIS structure functions
\begin{equation}
F_T=\frac{N^2Q^2x}{32\pi^3}\,\mathrm{Im}\,f_T(\alpha),\qquad
F_L=\frac{N^2Q^2x}{32\pi^3}\,\mathrm{Im}\,f_L(\alpha).
\label{eq:current-structure}
\end{equation}
Equation \eqref{eq:current-structure} holds in the strict near-lightcone limit \eqref{eq:limit}. The retarded prescription gives $\mathrm{Im}\,f_T>0$ and $\mathrm{Im}\,f_L>0$, as shown in Appendix~\ref{app:matching}, so both structure functions are positive. In Section~\ref{sec:currents}, we show how a Weyl interaction changes the argument of each of these two response functions.

The absorption exponent also has a direct near-lightcone form. Combining \eqref{eq:current-WKB-constant}, \eqref{eq:dict2} and \eqref{eq:dictlimit} gives
\begin{equation}
\begin{aligned}
\frac{A}{(128\rho)^{1/4}}&\longrightarrow A\alpha^{1/4}\\[5pt]
&=\frac{A}{2\pi}\left(\frac{xQ}{T}\right)^{1/2}.
\end{aligned}
\label{eq:expdict}
\end{equation}
The leading absorption is therefore proportional to $\exp(-A\alpha^{1/4})$ at large $\alpha$. The coefficient $A$, derived here from the two current equations, agrees with the leading WKB exponent for thermal stress-tensor correlators in Einstein gravity in Ref.~\cite{Esper:2023jeq}.

\section{Universality of the two current channels}
\label{sec:currents}

The calculation for a minimally coupled gauge field gives one response function for each polarization. We now ask how these functions change when the current couples differently to the stress tensor. In the bulk, this change is described by an interaction between the gauge field and the Weyl tensor. We will show that the interaction changes the argument of each response function while preserving its shape.

\subsection{Adding the Weyl interaction}

We  keep the Einstein black-brane metric \eqref{eq:metric} fixed. The gauge field is governed by
\begin{equation}
S=-\frac{1}{4g^2}\int \dd^5x\,\sqrt{-g}\,
\bigl(F_{MN}F^{MN}-4\gamma L^2 C_{MNPQ}F^{MN}F^{PQ}\bigr).
\label{eq:weyl-action}
\end{equation}
Here $M,N,P,Q$ are five-dimensional spacetime indices, and $g$ under the square root is the metric determinant. We keep the bulk gauge coupling $g^2$, and hence the current normalization \eqref{eq:current-normalization}, fixed while varying the dimensionless Weyl coupling $\gamma$. The AdS radius is $L$, $C_{MNPQ}$ is the Weyl tensor, and the field strength of the gauge potential $A_M$ is
\[
F_{MN}=\partial_M A_N-\partial_N A_M.
\]
Setting $\gamma=0$ recovers the minimally coupled gauge-field problem. The boundary field theory remains strongly coupled.

The boundary quantity controlled by this interaction is the parameter $\hat a_2$ in the current--current--stress-tensor three-point function~\cite{Hofman:2008ar,Hofman:2009ug,Ritz:2008kh}. One way to define $\hat a_2$ is to create a state with the current and measure its outgoing energy. Let $E_0$ be the total energy of the state and $\mathcal E(\theta)$ the mean energy per unit solid angle measured at angle $\theta$ to its real polarization vector. Then
\begin{equation}
\mathcal E(\theta)=\frac{E_0}{4\pi}\left[1+\hat a_2\left(\cos^2\theta-\frac13\right)\right],
\qquad -\frac32\leq \hat a_2\leq3.
\label{eq:current-collider}
\end{equation}
The bounds follow by requiring this energy to be nonnegative in every detector direction~\cite{Hofman:2008ar}. For the action \eqref{eq:weyl-action}, the two coupling conventions are related by
\begin{equation}
\gamma=-\frac{\hat a_2}{48}.
\label{eq:coupling-map}
\end{equation}

To find the thermal current response, keep the gauge and Fourier conventions of \eqref{eq:gauge-ansatz} and the two fields defined in \eqref{eq:current-masterfields}. The Weyl tensor modifies the coefficients of their radial equations through three functions:
\begin{equation}
h_T(u)=1+8\gamma u^2,\qquad h_L(u)=1-24\gamma u^2,\qquad h_S(u)=1-8\gamma u^2.
\label{eq:weyl-factors}
\end{equation}
In the equations below we suppress the radial argument $u$ of these functions. In the dimensionless momenta \eqref{eq:dimensionless-momenta}, the transverse and longitudinal equations are
\begin{align}
\frac{\dd}{\dd u}\left(fh_T\frac{\dd A_\perp}{\dd u}\right)
+\frac{w^2h_T-q^2fh_S}{uf}A_\perp&=0,
\label{eq:weyl-full-T}\\
\frac{\dd}{\dd u}\left[\frac{uf}{h_T}\frac{\dd(h_La)}{\dd u}\right]
+\left(\frac{w^2h_L}{fh_T}-q^2\right)a&=0.
\label{eq:weyl-full-L}
\end{align}
The first follows directly from the transverse gauge-field equation. The longitudinal components are coupled, but the radial constraint allows us to eliminate $A_z$ and obtain the second equation for $a$. Appendix~\ref{app:reduced} gives that elimination explicitly. When $\gamma=0$, all three coefficient functions equal one and the equations reduce to \eqref{eq:current-full}.

\subsection{The near-lightcone limit}

We take the limit \eqref{eq:epsilon-limit} of the full equations \eqref{eq:weyl-full-T} and \eqref{eq:weyl-full-L}, keeping the Weyl coupling $\gamma$ fixed. The corrections to $h_T$, $h_L$ and $h_S$ vanish as $u^2$, but the spatial momentum $q$ grows in this limit. We first show which terms survive in the transverse equation, then carry out the longitudinal reduction.

\subsubsection{Transverse polarization}

Dividing \eqref{eq:weyl-full-T} by $fh_T$ makes the coefficient of $A_\perp''$ equal to one. With primes denoting derivatives with respect to $u$, the equation becomes
\begin{equation}
A_\perp''+\left(\frac{f'}f+\frac{h_T'}{h_T}\right)A_\perp'
+\frac{w^2-q^2fh_S/h_T}{uf^2}A_\perp=0.
\label{eq:weyl-normalized-T}
\end{equation}
The functions in \eqref{eq:weyl-factors}, together with $f=1-u^2$, give
\begin{equation}
f\frac{h_S}{h_T}=1-r_Tu^2+O(u^4),\qquad
r_T=1+16\gamma=1-\frac{\hat a_2}{3},
\label{eq:thermal-factor-T}
\end{equation}
where the last equality uses \eqref{eq:coupling-map}. Since $K^2=q^2-w^2$, the numerator of the momentum term in \eqref{eq:weyl-normalized-T} is
\begin{equation}
w^2-q^2f\frac{h_S}{h_T}
=-K^2+r_Tq^2u^2+O(q^2u^4).
\label{eq:weyl-momentum-T}
\end{equation}
The virtuality contribution is $-K^2$, and the leading thermal contribution is $r_Tq^2u^2$. Thus the Weyl interaction changes the latter by the factor $r_T$.

The scaling \eqref{eq:epsilon-limit} gives
\begin{equation}
K^2=\frac{\alpha}{\epsilon^2},\qquad
q^2u^2=\frac{\alpha^2\ut^2}{4\epsilon^2}+O(\epsilon^2).
\label{eq:weyl-momentum-scaling}
\end{equation}
The virtuality term and the leading thermal term in \eqref{eq:weyl-momentum-T} are therefore of order $\epsilon^{-2}$. Dividing by $uf^2$ makes them of order $\epsilon^{-4}$, the same order as $A_\perp''$.

To obtain the equation in $\ut$, use $d/du=\epsilon^{-2}d/d\ut$ and multiply \eqref{eq:weyl-normalized-T} by $\epsilon^4$. At fixed $\ut$, its first-derivative coefficient and momentum coefficient become
\begin{equation}
\begin{aligned}
\epsilon^2\left(\frac{f'}f+\frac{h_T'}{h_T}\right)&=O(\epsilon^4),\\
\epsilon^4\frac{w^2-q^2fh_S/h_T}{uf^2}
&=-\frac{\alpha}{\ut}+r_T\frac{\alpha^2\ut}{4}+O(\epsilon^4).
\end{aligned}
\label{eq:weyl-scaled-T}
\end{equation}
The first derivative disappears as $\epsilon\to0$, while both momentum contributions remain. The transverse equation is
\begin{equation}
\frac{\dd^2 A_\perp}{\dd\ut^2}
=\left(\frac{\alpha}{\ut}-r_T\frac{\alpha^2\ut}{4}\right)A_\perp.
\label{eq:weyl-reduced-T}
\end{equation}

\subsubsection{Longitudinal polarization}

In \eqref{eq:weyl-full-L}, derivatives also act on $h_L$ inside $h_La$. Expanding those derivatives and dividing by the coefficient $ufh_L/h_T$ of $a''$ gives
\begin{equation}
\begin{aligned}
a''+\left(\frac1u+\frac{f'}f-\frac{h_T'}{h_T}
                 +2\frac{h_L'}{h_L}\right)a'
&+\frac{w^2-q^2fh_T/h_L}{uf^2}a\\
&+\left[\frac{h_L''}{h_L}
+\left(\frac1u+\frac{f'}f-\frac{h_T'}{h_T}\right)
\frac{h_L'}{h_L}\right]a=0.
\end{aligned}
\label{eq:weyl-normalized-L}
\end{equation}
The last bracket contains the extra terms from differentiating $h_L$. The momentum term has the same form as in \eqref{eq:weyl-normalized-T}, with $fh_S/h_T$ replaced by $fh_T/h_L$. Expanding this ratio gives
\begin{equation}
f\frac{h_T}{h_L}=1-r_Lu^2+O(u^4),\qquad
r_L=1-32\gamma=1+\frac{2\hat a_2}{3}.
\label{eq:thermal-factor-L}
\end{equation}
Equations \eqref{eq:weyl-momentum-T} and \eqref{eq:weyl-momentum-scaling} therefore apply to the longitudinal momentum term with $r_T$ replaced by $r_L$.

The coefficient of $a'$ and the extra bracket in \eqref{eq:weyl-normalized-L} behave as
\begin{equation}
\begin{aligned}
\frac1u+\frac{f'}f-\frac{h_T'}{h_T}+2\frac{h_L'}{h_L}
&=\frac1u+O(u),\\
\frac{h_L''}{h_L}
+\left(\frac1u+\frac{f'}f-\frac{h_T'}{h_T}\right)\frac{h_L'}{h_L}
&=O(1).
\end{aligned}
\label{eq:weyl-longitudinal-derivatives}
\end{equation}
After the change to $\ut$ and multiplication by $\epsilon^4$, the term $a'/u$ becomes $\ut^{-1}\,da/d\ut$ and survives. The $O(u)a'$ term and the $O(1)a$ term in \eqref{eq:weyl-longitudinal-derivatives} each acquire a factor $\epsilon^4$ and disappear. The resulting longitudinal equation is
\begin{equation}
\frac{\dd^2 a}{\dd\ut^2}+\frac1{\ut}\frac{\dd a}{\dd\ut}
=\left(\frac{\alpha}{\ut}-r_L\frac{\alpha^2\ut}{4}\right)a.
\label{eq:weyl-reduced-L}
\end{equation}

In both \eqref{eq:weyl-reduced-T} and \eqref{eq:weyl-reduced-L}, the omitted coefficient corrections are $O(\epsilon^4)$ at fixed positive $\ut$, $\alpha$ and $\gamma$. The Weyl interaction changes only the coefficient of the thermal term. The virtuality term $\alpha/\ut$ is unchanged, and the longitudinal first derivative remains.

The factors $r_T$ and $r_L$ in \eqref{eq:thermal-factor-T} and \eqref{eq:thermal-factor-L} are restricted by the conformal collider bounds \eqref{eq:current-collider}:
\begin{equation}
0\leq r_T\leq\frac32,\qquad
0\leq r_L\leq3,\qquad
2r_T+r_L=3.
\label{eq:current-ri-bounds}
\end{equation}
At $\hat a_2=3$ the transverse factor $r_T$ vanishes; at $\hat a_2=-3/2$ the longitudinal factor $r_L$ vanishes. The corresponding reduced equation then contains no thermal term.

\subsection{Kaigorodov geometry}
\label{sec:kaigorodov}

The same reduced equations can be obtained by taking the limit of the geometry before varying the gauge-field action. 
This gives an alternative explanation of why the Weyl interaction changes only the two thermal couplings.
 Let $t$ be physical time and $z$ the physical boundary coordinate along the spatial momentum $k$. Define dimensionless lightcone coordinates $x^\pm$ and their conjugate momenta $p_\pm$ by
\begin{align}
x^\pm&=\pi T(t\pm z),\qquad
-\omega t+kz=p_+x^++p_-x^-,\nonumber\\
p_+&=\frac{k-\omega}{2\pi T}>0,\qquad
p_-=-\frac{k+\omega}{2\pi T}<0.
\label{eq:lc-momenta}
\end{align}
Thus $K^2=-p_+p_-$ and $\alpha=-p_-p_+^3$. Write $\mathbf x_\perp=\pi T\mathbf z_\perp$ for the two dimensionless transverse coordinates, where $\mathbf z_\perp$ are the physical ones. The position-space form of the limit is the scaling $x^+\sim\epsilon^{-1}$, $x^-\sim\epsilon^3$, $u\sim\epsilon^2$ and $\mathbf x_\perp\sim\epsilon$. Keeping the terms of \eqref{eq:metric} that survive this scaling and writing $\zeta=\sqrt u$ gives
\begin{equation}
\frac{ds_5^2}{L^2}=\frac{1}{\zeta^2}\Big[-dx^+dx^-+d\zeta^2+d\mathbf x_\perp^2+\frac{\zeta^4}{4}(dx^+)^2\Big].
\label{eq:kaigorodov}
\end{equation}
Here and below we use the same coordinate and momentum symbols for the limiting geometry, with $K^2=-p_+p_-$. The combinations $\alpha=-p_-p_+^3$ and $v=-p_+p_-u$ are unchanged by the scaling; Appendix~\ref{app:reduced} gives the coordinate and field transformations explicitly.

The metric \eqref{eq:kaigorodov} is the Kaigorodov spacetime~\cite{Cvetic:1998jf,Brecher:2000pa}. It is an exact solution of Einstein's equations with negative cosmological constant, and can also be obtained by boosting a black brane to infinite momentum while sending its temperature to zero at fixed momentum density~\cite{Cvetic:1998jf,Brecher:2000pa,Singh:2010zs,Armas:2026mul}. Its dual state has a single nonzero stress-tensor component, $\langle T_{++}\rangle$.

The difference from anti-de Sitter space is the single $(dx^+)^2$ term. In the inverse metric it changes only $g^{--}=-\zeta^6/L^2$, while the volume element remains $\sqrt{-g}=L^5/(2\zeta^5)$. Consequently the thermal part of $F_{MN}F^{MN}$ involves only $F_{-\zeta}^2$ and $F_{-\perp}^2$, where $F_{-\perp}^2=F_{-1}^2+F_{-2}^2$ and the indices $1,2$ denote the two transverse coordinates. The longitudinal sector contains $F_{-\zeta}$; a transverse Fourier mode at zero transverse momentum contains $F_{-\perp}$.

The Weyl tensor has only the components $C_{+a+b}$ and those related by its index symmetries, where $a,b$ range over $\zeta,x_1,x_2$. Its nonzero independent entries are
\begin{equation}
C_{+\zeta+\zeta}=-L^2,\qquad
C_{+1+1}=C_{+2+2}=\frac{L^2}{2}.
\label{eq:kaigorodov-weyl}
\end{equation}
Since $g^{++}=0$ and $g^{+-}=-2\zeta^2/L^2$, raising the indices gives $F^{+a}=-2\zeta^4 F_{-a}/L^4$. The Weyl interaction therefore involves exactly the same field-strength components and radial power as the thermal part of the minimal action. The full contraction gives
\begin{equation}
\begin{aligned}
F_{MN}F^{MN}-4\gamma L^2C_{MNPQ}F^{MN}F^{PQ}
={}&\big(F_{MN}F^{MN}\big)_{\rm AdS}\\
&-\frac{2\zeta^8}{L^4}\Big[r_TF_{-\perp}^2+r_LF_{-\zeta}^2\Big].
\end{aligned}
\label{eq:wave-vertex}
\end{equation}
The first term is evaluated with the anti-de Sitter metric and the same covariant field strength. The identity holds for an arbitrary field strength, with no gauge choice or expansion in $\gamma$. Together with the unchanged volume element, it shows that the Weyl interaction only multiplies the thermal term by $r_T$ or $r_L$ in the respective channel. The factors are those already found in \eqref{eq:thermal-factor-T} and \eqref{eq:thermal-factor-L}; no new radial coupling is generated for the action \eqref{eq:weyl-action}.

To obtain the channel equations, take radial gauge $A_\zeta=0$ and Fourier dependence $e^{i(p_+x^++p_-x^-)}$, with zero transverse momentum. Let $\mathcal A(\zeta)$ be a transverse gauge-potential amplitude, and let $A_+$ and $A_-$ be the amplitudes multiplying $dx^+$ and $dx^-$ in the gauge one-form. The longitudinal variable in the limiting geometry is $\psi=dA_-/du$, where $u=\zeta^2$. The two radial equations following directly from \eqref{eq:wave-vertex} are
\begin{align}
\mathcal A''-\frac{\mathcal A'}{\zeta}
+\Big[4p_+p_-+r_Tp_-^2\zeta^4\Big]\mathcal A&=0,
\label{eq:maxwell-kaigorodov}\\
\psi''+\frac{\psi'}{\zeta}
+\Big[4p_+p_-+r_Lp_-^2\zeta^4\Big]\psi&=0,
\label{eq:longitudinal-kaigorodov}
\end{align}
where primes in these two equations denote $\zeta$ derivatives. The longitudinal equation uses the radial constraint to eliminate $A_+$; Appendix~\ref{app:reduced} gives this step and shows that $\psi$ is the normalized near-lightcone limit of $a=dA_t/du$, with the same ratio of boundary coefficients.

Using $\tilde u=\zeta^2/p_+^2$ and $\alpha=-p_-p_+^3$, these two equations become \eqref{eq:weyl-reduced-T} and \eqref{eq:weyl-reduced-L}, respectively. The geometry thus explains the channel reductions already obtained from the black-brane equations. To determine the response one must also specify the interior condition. We use the incoming condition inherited from the retarded black-brane problem.

\subsection{From the reduced equations to the response functions}

We next show that the factors $r_T$ and $r_L$ can be absorbed into the argument of the corresponding response at $\gamma=0$. Use the radial coordinate $v$ of \eqref{eq:fixed-v}, which also appears in the boundary expansions. The equations become
\begin{align}
\frac{\dd^2 A_\perp}{\dd v^2}
&=\left(\frac1v-\frac{r_Tv}{4\alpha}\right)A_\perp,
\label{eq:universal-T}\\
\frac{\dd^2a}{\dd v^2}+\frac1v\frac{\dd a}{\dd v}
&=\left(\frac1v-\frac{r_Lv}{4\alpha}\right)a.
\label{eq:universal-L}
\end{align}
The longitudinal field is still $a=\dd A_t/\dd u$, now evaluated as a function of $v$.

Each equation depends on the Weyl coupling and on $\alpha$ only through the ratio $r_i/\alpha$. For a channel with $r_i>0$, define
\begin{equation}
\alpha_i=\frac{\alpha}{r_i},\qquad i=T,L.
\label{eq:alpha-i}
\end{equation}
At fixed $v$, the equation is then exactly the corresponding equation at $\gamma=0$, evaluated at $\alpha_i$. The boundary remains at $v=0$, and the incoming-wave prescription at large $v$ is the same. The differential equation and the prescription selecting its solution therefore agree in the two problems.

The response is read from the ratio of boundary coefficients, with the sign convention in \eqref{eq:current-boundary}. Because $v$ is unchanged, the source terms and the logarithm $\log v$ used to define those coefficients are unchanged too. The same solution with the same boundary convention gives the same response. Let $f_T^{(0)}$ and $f_L^{(0)}$ denote the functions at vanishing Weyl coupling. We obtain
\begin{equation}
\begin{aligned}
f_T(\alpha;\gamma)&=f_T^{(0)}(\alpha_T),\\
f_L(\alpha;\gamma)&=f_L^{(0)}(\alpha_L).
\end{aligned}
\label{eq:response-universality}
\end{equation}
This is the universality statement: changing the Weyl coupling changes the argument of each channel's function. Each polarization has its own response function at $\gamma=0$. Keeping $v$ fixed also makes clear why no additional $\log r_i$ shift appears in the boundary coefficients.

There is one further step before applying this statement to a current correlator. The radial solution determines a ratio of coefficients, but the overall normalization of the correlator comes from the on-shell action. Two identical radial equations could still give differently normalized correlators if their boundary actions differed by a finite factor.

Returning to the original black-brane coordinates of \eqref{eq:metric}, the transverse and temporal boundary terms in the on-shell action contain the weights
\begin{equation}
\begin{aligned}
\sqrt{-g}\,g^{uu}g^{\perp\perp}h_T
&=\frac{2r_+^2}{L^3}f(1+8\gamma u^2),\\
\sqrt{-g}\,g^{uu}g^{tt}h_L
&=-\frac{2r_+^2}{L^3}(1-24\gamma u^2).
\end{aligned}
\label{eq:boundary-factors}
\end{equation}
Both weights have finite limits as $u\to0$, and their explicit Weyl corrections vanish. The spatial longitudinal boundary weight has the same limiting behavior as the transverse one. With the current normalization \eqref{eq:current-normalization} held fixed, the relation between the boundary responses and the structure functions in \eqref{eq:current-structure} is unchanged. All dependence on $\gamma$ is therefore contained in the functions in \eqref{eq:response-universality}.

\subsection{Freedom near the lightcone: the vacuum DIS response}
\label{sec:current-freedom}

At a conformal collider endpoint, the selected current channel has the vacuum response in the near-lightcone limit. 
Since a spacelike current cannot be absorbed by the vacuum, its normalized DIS absorption vanishes in this limit. 
We now explain this in more detail.

The corresponding phenomenon for stress-tensor correlators was established in  Ref.~\cite{Huang:2022vet}. 
In that work, saturation of the appropriate averaged null energy condition (ANEC) removes the thermal dependence from the leading near-lightcone equations in Gauss--Bonnet gravity. 
The selected thermal stress-tensor correlator then takes its vacuum form. 
Here we establish the current version of this statement in the model with the gauge-field--Weyl interaction.

The mechanism is visible directly in \eqref{eq:universal-T} and \eqref{eq:universal-L}. The factors $r_T$ and $r_L$ multiply the thermal terms. They also multiply the energy flux in \eqref{eq:current-collider} for a detector perpendicular and parallel to the current polarization, respectively. At the two endpoints,
\begin{equation}
\begin{aligned}
\hat a_2=3 &: \qquad r_T=0,\\
\hat a_2=-\frac32 &: \qquad r_L=0.
\end{aligned}
\label{eq:current-endpoints}
\end{equation}
Setting the appropriate factor to zero reduces that channel's radial equation to the vacuum form.

For spacelike momentum, the vacuum prescription selects the solution that decays in the interior. In the transverse channel at $\hat a_2=3$ and the longitudinal channel at $\hat a_2=-3/2$, these solutions are, respectively,
\begin{equation}
\begin{aligned}
\frac{A_\perp(v)}{C_T}&=2\sqrt v\,K_1(2\sqrt v),\\
\frac{a(v)}{C_L}&=-2K_0(2\sqrt v),
\end{aligned}
\label{eq:current-endpoint-profiles}
\end{equation}
where $K_\nu$ is the modified Bessel function of the second kind of order $\nu$. The source coefficients $C_T$ and $C_L$ are those defined in Section~\ref{sec:nearlc}. Expanding at $v=0$ and comparing with \eqref{eq:current-boundary} gives
\begin{equation}
\begin{aligned}
f_T\big|_{\hat a_2=3}&=2\gamma_E-1,\\
f_L\big|_{\hat a_2=-3/2}&=-2\gamma_E,
\end{aligned}
\label{eq:current-endpoint-responses}
\end{equation}
where $\gamma_E$ is Euler's constant. These are precisely the real vacuum response functions, in the same subtraction convention. Together with the unchanged boundary normalization, they give the vacuum current response in the selected channel.

In particular, the imaginary parts vanish. Using \eqref{eq:current-structure}, the DIS structure functions at leading order in the near-lightcone limit satisfy
\begin{equation}
\begin{aligned}
F_T\big|_{\hat a_2=3}&=0,\\
F_L\big|_{\hat a_2=-3/2}&=0.
\end{aligned}
\label{eq:current-endpoint-DIS}
\end{equation}
Here the near-lightcone limit is taken at fixed $\alpha$ as in \eqref{eq:limit}. The real vacuum polarization remains. Only the indicated polarization decouples at each endpoint; the other polarization retains a thermal response. Subleading corrections away from the near-lightcone limit can also retain temperature dependence, as in Ref.~\cite{Huang:2022vet}.

\subsection{Numerical results}
\label{sec:current-numerics}

We compute $f_T^{(0)}(\alpha)$ and $f_L^{(0)}(\alpha)$ by solving the reduced equations \eqref{eq:universal-T} and \eqref{eq:universal-L} at vanishing Weyl coupling, $\gamma=0$, with incoming boundary conditions. Matching to the boundary expansions \eqref{eq:current-boundary} determines the two functions. For $r_i>0$, the functions $f_T$ and $f_L$ at nonzero $\hat a_2$ then follow by rescaling the argument according to \eqref{eq:response-universality}. Appendix~\ref{app:numerics} describes the numerical methods and independent checks.

The quantities relevant to DIS are $\Im f_T$ and $\Im f_L$: multiplying them by the common positive prefactor in \eqref{eq:current-structure} gives $F_T$ and $F_L$, respectively. With the boundary definitions \eqref{eq:current-boundary}, the incoming radial flux is proportional to $\Im f_T$ for $A_\perp$ and to $\Im f_L$ for $a$, as shown in Appendix~\ref{app:matching}. The positive absorption agrees with the longitudinal and transverse solutions in Eqs.~(4.7) and (4.10) of Ref.~\cite{Hatta:2007cs}.

\begin{figure}[!htbp]
\centering
\includegraphics[width=\textwidth]{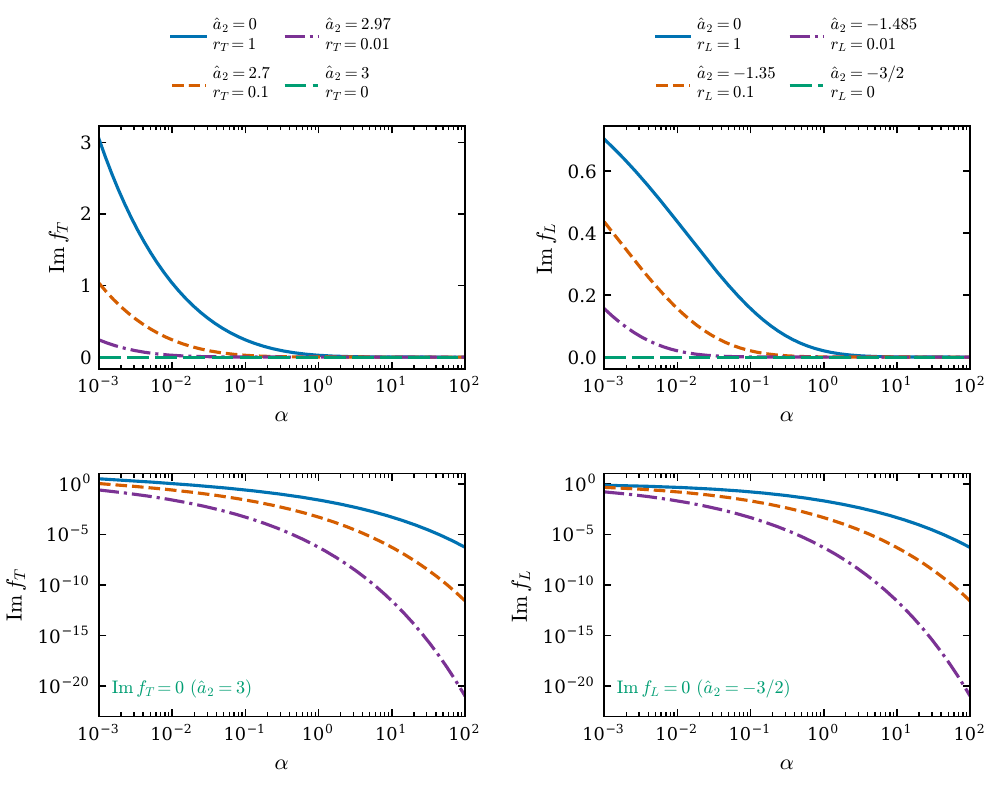}
\caption{Absorption near conformal collider saturation. The transverse panels show $\Im f_T$ at $\hat a_2=0,2.7,2.97,3$; the longitudinal panels show $\Im f_L$ at $\hat a_2=0,-1.35,-1.485,-1.5$. In each channel these choices give $r_i=1,0.1,0.01,0$. All horizontal axes show $\alpha$ on a logarithmic scale. The upper panels have linear vertical axes and include the curves at zero for $r_i=0$. The lower panels have logarithmic vertical axes, so these zero curves are omitted. Each nonzero curve is obtained from $f_T^{(0)}$ or $f_L^{(0)}$ using \eqref{eq:response-universality}.}
\label{fig:current-responses-absorption}
\end{figure}

To show how absorption vanishes at the conformal collider endpoints, we choose $r_i=1,0.1,0.01,0$. For the transverse channel these correspond to $\hat a_2=0,2.7,2.97,3$; for the longitudinal channel they correspond to $\hat a_2=0,-1.35,-1.485,-1.5$. Figure~\ref{fig:current-responses-absorption} shows $\Im f_T$ and $\Im f_L$ for these values. The upper panels use linear vertical axes and include the horizontal curves at zero for $r_i=0$. The lower panels use logarithmic vertical axes to resolve the exponentially small absorption at positive $r_i$.

For $r_i>0$, the rescaling in \eqref{eq:response-universality} shifts each curve horizontally on a logarithmic $\alpha$ axis. The plots show the crossover from strong to weak thermal absorption as $\alpha/r_i$ increases. At fixed $\alpha$, decreasing $r_i$ increases the argument $\alpha_i=\alpha/r_i$ in \eqref{eq:alpha-i} and suppresses absorption. As $r_i\to0^+$, the WKB suppression factor is $\exp[-A(\alpha/r_i)^{1/4}]$, with $A$ defined in \eqref{eq:current-WKB-constant}. At $r_i=0$, \eqref{eq:current-endpoint-DIS} gives zero absorption in that channel.

\begin{figure}[!htbp]
\centering
\includegraphics[width=\textwidth]{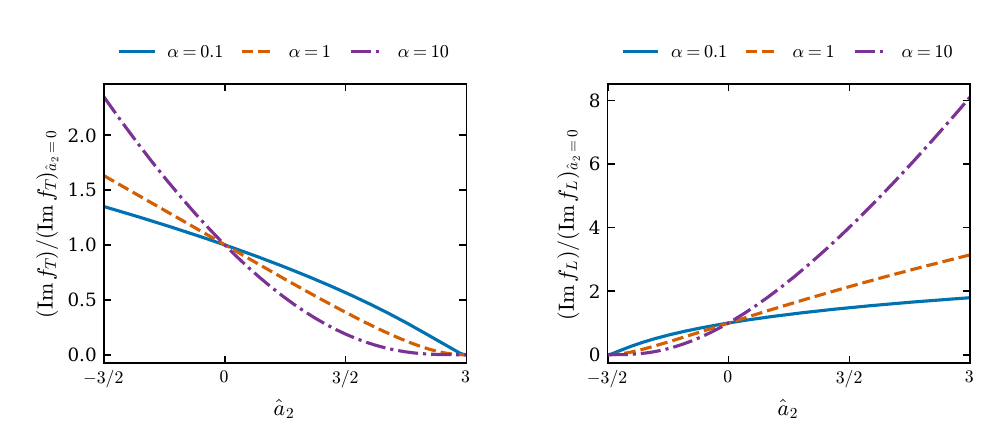}
\caption{$\Im f_T$ and $\Im f_L$ as functions of $\hat a_2$ at fixed $\alpha=0.1,1,10$, each divided by its value at $\hat a_2=0$ and the same $\alpha$. Both axes are linear. The exact endpoint values are included: $\Im f_T=0$ at $\hat a_2=3$, and $\Im f_L=0$ at $\hat a_2=-3/2$, as in \eqref{eq:current-endpoint-DIS}. For $r_i>0$, the curves are obtained from $f_T^{(0)}$ and $f_L^{(0)}$ using \eqref{eq:response-universality}.}
\label{fig:current-responses-coupling}
\end{figure}

Figure~\ref{fig:current-responses-coupling} shows $\Im f_T$ and $\Im f_L$ as functions of $\hat a_2$ at fixed $\alpha=0.1,1,10$, each divided by its own value at $\hat a_2=0$ and the same $\alpha$. Transverse absorption decreases to zero as $\hat a_2\to3$, while longitudinal absorption decreases to zero as $\hat a_2\to-3/2$. At either endpoint, absorption vanishes only in the indicated channel.

\begin{figure}[!htbp]
\centering
\includegraphics[width=\textwidth]{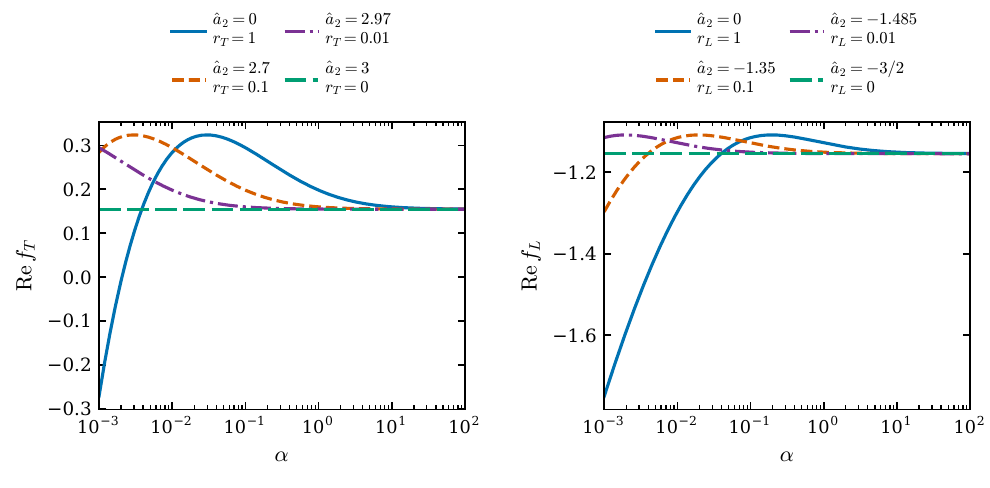}
\caption{Real parts of $f_T$ and $f_L$ near conformal collider saturation, with the same channel-dependent values of $\hat a_2$ as in Figure~\ref{fig:current-responses-absorption}. The horizontal axes show $\alpha$ on a logarithmic scale. The constant curves at $\hat a_2=3$ in the transverse channel and $\hat a_2=-3/2$ in the longitudinal channel give the exact vacuum values in \eqref{eq:current-endpoint-responses}.}
\label{fig:current-responses-real}
\end{figure}

Figure~\ref{fig:current-responses-real} shows $\Re f_T$ and $\Re f_L$ for the same choices of $r_i$ as the absorption curves. At large $\alpha$, they approach the vacuum constants in \eqref{eq:current-endpoint-responses}, with thermal corrections obtained from the expansion in Appendix~\ref{app:moments} and the rescaling \eqref{eq:response-universality}. At $r_i=0$, the corresponding real part equals its vacuum constant for all $\alpha$.

\begin{samepage}
At $\gamma=0$, Eqs.~(4.7) and (4.10)--(4.13) of Ref.~\cite{Hatta:2007cs} give
\begin{equation}
\mathrm{Im}\,f_T^{(0)}(\alpha)\sim
\frac{3\pi}{6^{2/3}\Gamma(1/3)^2}\,\alpha^{-1/3},
\qquad
\mathrm{Im}\,f_L^{(0)}(\alpha)\longrightarrow\frac\pi3,
\qquad \alpha\to0,
\label{eq:current-small-alpha-absorption}
\end{equation}
where $\Gamma$ is the Euler gamma function. The numerical results agree with these limits. At large $\alpha$, $\Im f_T^{(0)}$ and $\Im f_L^{(0)}$ exhibit the WKB suppression determined by \eqref{eq:expdict}. The real parts agree with the thermal expansion derived in Appendix~\ref{app:moments} at $\gamma=0$.
\end{samepage}

\clearpage

\section{The absorptive exponent and the complex null geodesic}
\label{sec:geodesic}

$\Im f_T$ and $\Im f_L$ are exponentially small when the bulk gauge field tunnels through a barrier. We first calculate their WKB exponent. We then construct a complex null geodesic of the Kaigorodov geometry \eqref{eq:kaigorodov} from the metric alone. It is the curve defined by the gradient of the WKB phase, and the eikonal phase $\delta$ computed from its endpoint separations $\Delta x^\pm$ reproduces this exponent. From the same separations we compute the invariant $\mathcal X=-16\Delta x^-(\Delta x^+)^3$ and compare it with the scalar stress-sector singularity location $\mathcal X_c$ found in Ref.~\cite{Araya:2026shz}.

In this section we set the AdS radius $L=1$ and the Weyl coupling $\gamma=0$. First we take the near-lightcone limit \eqref{eq:limit}, keeping the positive momentum parameter $\alpha$ fixed. We then take $\alpha\gg1$ to justify WKB. The rescalings for nonzero $\gamma$ are given in \eqref{eq:alpha-i} and \eqref{eq:response-universality}.

\subsection{Tunnelling through the barrier}
\label{sec:wave-phase}

The WKB calculation follows the same reasoning as the stress-tensor calculation in Ref.~\cite{Esper:2023jeq}. Recall the parameter $\alpha=K^6/(q^+)^2$ in \eqref{eq:dict1}, where $K^2=(k^2-\omega^2)/(2\pi T)^2$ and $q^+=(\omega+k)/(2\pi T)$. Here $\omega$ is the frequency, $k>\omega>0$ the spatial momentum, and $T$ the temperature.

We use the radial coordinate $s$, defined in terms of $v=K^2u$ from \eqref{eq:fixed-v} by
\begin{equation}
s=\frac{v}{2\sqrt\alpha}=\frac{K^2u}{2\sqrt\alpha}.
\label{eq:barrier-coordinate}
\end{equation}
The boundary is at $s=0$ and the turning point at $s=1$, independent of $\alpha$.

\subsubsection{Transverse polarization}

For the transverse gauge-field component $A_\perp$, the radial field $\Phi_T=A_\perp$ satisfies
\begin{equation}
\frac{d^2\Phi_T}{ds^2}
=2\sqrt\alpha\left(\frac1s-s\right)\Phi_T.
\label{eq:transverse-s}
\end{equation}
Equation~\eqref{eq:transverse-s} follows from the first equation in \eqref{eq:reduced}, using $v=\alpha\tilde u$ from \eqref{eq:v-lightcone} and the change of variable \eqref{eq:barrier-coordinate}. The potential $2\sqrt\alpha(1/s-s)$ is positive for $0<s<1$, where the solutions are exponential, and negative for $s>1$, where they are oscillatory.

For the WKB approximation, we write the transverse field $\Phi_T(s)$ as a slowly varying amplitude $\mathcal A_T(s)$ times the phase factor $e^{iW(s)}$. The radial momentum $p_s$ is the derivative of the radial phase $W$:
\begin{equation}
\Phi_T(s)\simeq\mathcal A_T(s)e^{iW(s)},
\qquad p_s=\frac{dW}{ds}.
\label{eq:wkb-phase}
\end{equation}
At leading WKB order, we neglect $dp_s/ds$ compared with $p_s^2$, together with derivatives of the amplitude $\mathcal A_T$. Substituting \eqref{eq:wkb-phase} into \eqref{eq:transverse-s} gives
\begin{equation}
p_s^2=-2\sqrt\alpha\left(\frac1s-s\right).
\label{eq:ps}
\end{equation}
Inside the barrier, the radial momentum $p_s$ is imaginary. Choosing $p_s=+i|p_s|$ gives a transverse field $\Phi_T$ that decreases as $s$ increases.

WKB requires the radial momentum $p_s$ to change little over the local decay length $1/|p_s|$. For $p_s$ in \eqref{eq:ps}, this condition is
\begin{equation}
\frac{|dp_s/ds|}{|p_s|^2}
=\frac{1+s^2}{2\sqrt2\,\alpha^{1/4}\sqrt{s}\,(1-s^2)^{3/2}}
\ll1,
\qquad 0<s<1.
\label{eq:wkb-validity-scaled}
\end{equation}
At fixed $0<s<1$, the ratio in \eqref{eq:wkb-validity-scaled} decreases as $\alpha^{-1/4}$. Thus WKB requires large $\alpha$ in addition to the near-lightcone limit \eqref{eq:limit}. The ratio diverges at the boundary $s=0$ and at the turning point $s=1$. In those regions, separate local solutions must be matched to the WKB solution \eqref{eq:wkb-phase}; Appendix~\ref{app:matching} gives this matching.

\subsubsection{Longitudinal polarization}

The longitudinal variable $a=dA_t/du$ is defined in \eqref{eq:current-masterfields}, where $A_t$ is the time component of the gauge field. The redefinition $\Phi_L=\sqrt v\,a$ removes the first derivative from the second equation in \eqref{eq:reduced}. Using $v=\alpha\tilde u$ from \eqref{eq:v-lightcone} and the change of variable \eqref{eq:barrier-coordinate}, the longitudinal field $\Phi_L$ satisfies
\begin{equation}
\frac{d^2\Phi_L}{ds^2}
=\left[2\sqrt\alpha\left(\frac1s-s\right)-\frac1{4s^2}\right]\Phi_L.
\label{eq:longitudinal-s}
\end{equation}
The term $2\sqrt\alpha(1/s-s)$ is the same as in the transverse equation \eqref{eq:transverse-s}. The extra term $-1/(4s^2)$ comes from the redefinition $\Phi_L=\sqrt v\,a$. Their relative magnitude in the barrier is
\begin{equation}
\frac{1/(4s^2)}{2\sqrt\alpha(1/s-s)}
=\frac1{8\sqrt\alpha\,s(1-s^2)}.
\label{eq:longitudinal-relative}
\end{equation}
At fixed $0<s<1$, the ratio in \eqref{eq:longitudinal-relative} tends to zero as $\alpha\to\infty$. Hence the fields $\Phi_T$ and $\Phi_L$ have the same leading radial momentum $p_s$ in \eqref{eq:ps}. This comparison is not uniform near $s=0$ or $s=1$. In particular, $\Phi_T$ and $\Phi_L$ have different boundary behavior. The matching in Appendix~\ref{app:matching} establishes that both fields nevertheless have the same leading absorption exponent.

\subsubsection{The WKB absorption exponent}

Define the positive barrier integral $I(\alpha)$ using the radial momentum $p_s$ in \eqref{eq:ps}:
\begin{equation}
I(\alpha)=\int_0^1 |p_s(s)|\,ds
=\sqrt2\,\alpha^{1/4}\int_0^1\sqrt{\frac1s-s}\,ds.
\label{eq:one-crossing}
\end{equation}
The phase factor $e^{iW}$ in \eqref{eq:wkb-phase} gives an attenuation $e^{-I}$ for one crossing of the barrier. The matching in Appendix~\ref{app:matching} justifies extending the integral \eqref{eq:one-crossing} to $s=0$ and $s=1$, although the WKB solution itself is not valid at those endpoints.

At fixed frequency $\omega$ and momentum $k$, the conserved radial flux $\mathcal J_T$ is proportional to the energy flux through the horizon and hence to the energy absorption rate in the plasma. We compare solutions with the boundary source held fixed. Let $C$ denote the normalization of the incoming wave. The matching explained in Appendix~\ref{app:matching} shows that its boundary source $C_T$, defined in \eqref{eq:current-boundary}, grows as $Ce^I$ at leading exponential order. Keeping $C_T$ fixed therefore requires $C\propto e^{-I}$. The incoming radial flux $\mathcal J_T$ is proportional to $|C|^2$, and is consequently suppressed by $e^{-2I}$. Appendix~\ref{app:matching} also derives the relation between $\mathcal J_T$ and $\Im f_T$, and the corresponding relation between the longitudinal flux $\mathcal J_L$ and $\Im f_L$. Thus $\Im f_T$ and $\Im f_L$ inherit the suppression $e^{-2I}$ when their boundary sources are held fixed.

The retarded condition requires a particular combination of the growing and decreasing exponentials under the barrier, derived in Appendix~\ref{app:matching}. Choosing only $p_s=+i|p_s|$ identifies the local decay rate but does not impose that boundary condition.

Evaluating \eqref{eq:one-crossing} gives
\begin{equation}
2I(\alpha)=A\alpha^{1/4},
\qquad
A=\frac{2\sqrt{2\pi}\,\Gamma(1/4)}{3\Gamma(3/4)}
=4.9442\ldots .
\label{eq:geodesic-action-value}
\end{equation}
Here $\Gamma$ is the Euler gamma function. The suppression factor agrees with the DIS result of Ref.~\cite{Hatta:2007cs} and the stress-tensor result of Ref.~\cite{Esper:2023jeq}.

$\Im f_T$ and $\Im f_L$ therefore obey
\begin{align}
\log\Im f_T(\alpha)&=-A\alpha^{1/4}+o(\alpha^{1/4}),\nonumber\\
\log\Im f_L(\alpha)&=-A\alpha^{1/4}+o(\alpha^{1/4}).
\label{eq:retarded-rate}
\end{align}
For the incoming solution, the flux argument in Appendix~\ref{app:matching} gives $\Im f_T>0$ and $\Im f_L>0$. The terms denoted by $o(\alpha^{1/4})$ are smaller than $\alpha^{1/4}$ at large $\alpha$.

\subsection{Complex null geodesics}
\label{sec:reduced-metric}
\label{sec:eikonal}

Consider null geodesics of the Kaigorodov metric in Section~\ref{sec:kaigorodov}. We use the lightcone coordinates \eqref{eq:lc-momenta} and set the transverse momentum $\mathbf p_\perp$, conjugate to $\mathbf x_\perp$, to zero.\footnote{At coincident transverse endpoints, the phase is stationary at zero transverse momentum. The Gaussian integral over transverse momentum contributes to the prefactor of the exponential.} The transverse coordinates are then constant, and the metric restricts to
\begin{equation}
ds_3^2
=\frac{u}{4}(dx^+)^2-\frac{dx^+dx^-}{u}
+\frac{du^2}{4u^2},
\label{eq:reduced-metric}
\end{equation}
where $u=\zeta^2$. This is the planar near-lightcone metric of Ref.~\cite{Parnachev:2020fna}. In this subsection we construct its geodesic directly from the metric.

A dot denotes differentiation with respect to a real affine parameter. Let $g_{ab}$ be the metric \eqref{eq:reduced-metric}, with indices $a,b$ running over $x^+,x^-,u$. The momenta are
\begin{equation}
\begin{aligned}
g_{+a}\dot x^a&=\frac u4\,\dot x^+-\frac{\dot x^-}{2u},\\[5pt]
g_{-a}\dot x^a&=-\frac{\dot x^+}{2u},\\[5pt]
g_{ua}\dot x^a&=\frac{\dot u}{4u^2}.
\end{aligned}
\label{eq:conserved-momenta}
\end{equation}
The first two are conserved because the metric is independent of $x^\pm$. For real conserved momenta $g_{\pm a}\dot x^a=p_\pm$, the null condition $g_{ab}\dot x^a\dot x^b=0$ gives
\begin{equation}
\left(\frac{\dot u}{4u^2}\right)^2
=\frac{p_+p_-}{u}+\frac{p_-^2u}{4}.
\label{eq:real-geodesic}
\end{equation}
If $p_+p_-<0$, the right-hand side is negative near $u=0$. If $p_+p_-\ge0$, it has no positive turning point. Thus no real null geodesic with $u>0$ leaves the boundary and returns to it, as for the black brane~\cite{Araya:2026shz}.

A returning complex geodesic is obtained by keeping $u$ and the affine parameter real and taking imaginary conserved momenta. Writing these in terms of real parameters $p_+>0>p_-$, we set
\begin{equation}
\begin{aligned}
g_{+a}\dot x^a&=-ip_+,\\[4pt]
g_{-a}\dot x^a&=-ip_-.
\end{aligned}
\label{eq:imaginary-momenta}
\end{equation}
The null condition becomes
\begin{equation}
\begin{aligned}
\left(\frac{\dot u}{4u^2}\right)^2
&=-\frac{p_+p_-}{u}-\frac{p_-^2u}{4}\\[5pt]
&=\frac{p_+(-p_-)}{u}\left(1-\frac{u^2}{u_*^2}\right),
\end{aligned}
\label{eq:complex-geodesic}
\end{equation}
with turning point
\begin{equation}
u_*=2\sqrt{\frac{p_+}{-p_-}}.
\label{eq:geodesic-turning-point}
\end{equation}
The right-hand side of \eqref{eq:complex-geodesic} is positive for $0<u<u_*$. The geodesic leaves the boundary, reaches $u_*$ and returns. Its boundary components follow from \eqref{eq:conserved-momenta} and \eqref{eq:imaginary-momenta}:
\begin{equation}
\begin{aligned}
\dot x^+&=2iu\,p_-,\\[4pt]
\dot x^-&=i\left(2u\,p_++u^3p_-\right).
\end{aligned}
\label{eq:affine-momenta}
\end{equation}
Thus $u$ is real, while $x^\pm$ change by imaginary amounts. By \eqref{eq:lc-momenta}, so do the boundary time $t$ and spatial coordinate $z$.

\begin{samepage}
Let $\Delta x^\pm$ denote the final minus initial boundary coordinates. Dividing \eqref{eq:affine-momenta} by the outgoing root $\dot u>0$ of \eqref{eq:complex-geodesic} gives $dx^\pm/du$. The returning segment contributes equally, so
\begin{equation}
\begin{aligned}
\Delta x^+
&=2\int_0^{u_*}\frac{\dot x^+}{\dot u}\,du\\[5pt]
&=-i\sqrt{\frac{-p_-}{p_+}}\,\sqrt{u_*}\,
\frac{\sqrt\pi\,\Gamma(1/4)}{2\Gamma(3/4)}.
\end{aligned}
\label{eq:plus-endpoint-integral}
\end{equation}
\end{samepage}
Using $u_*$ from \eqref{eq:geodesic-turning-point} and $A$ from \eqref{eq:geodesic-action-value} gives
\begin{equation}
\Delta x^+=-i\frac{3A}{4}
\left(\frac{-p_-}{p_+}\right)^{1/4}.
\label{eq:positive-plus}
\end{equation}
Similarly,
\begin{equation}
\begin{aligned}
\Delta x^-
&=2\int_0^{u_*}\frac{\dot x^-}{\dot u}\,du\\[5pt]
&=\frac{i}{3}\sqrt{\frac{p_+}{-p_-}}\,\sqrt{u_*}\,
\frac{\sqrt\pi\,\Gamma(1/4)}{2\Gamma(3/4)},
\end{aligned}
\label{eq:minus-endpoint-integral}
\end{equation}
and therefore
\begin{equation}
\Delta x^-=\frac{iA}{4}
\left(\frac{p_+}{-p_-}\right)^{3/4}.
\label{eq:positive-minus}
\end{equation}

\subsection{The geodesic from the WKB phase}
\label{sec:geodesic-wkb}

We now derive the same geodesic from the WKB solution of Section~\ref{sec:wave-phase}. Including the boundary Fourier factor in \eqref{eq:lc-momenta}, its total phase is
\begin{equation}
\mathcal S=p_+x^++p_-x^-+W(u).
\label{eq:eikonal-phase}
\end{equation}
Here $p_\pm$ are the real Fourier momenta of the gauge field, and $W$ is the radial phase in \eqref{eq:wkb-phase}. Changing from $s$ to $u$ with \eqref{eq:barrier-coordinate}, and using $K^2=-p_+p_-$ and $\alpha=-p_-p_+^3$, the WKB equation \eqref{eq:ps} gives
\begin{equation}
\begin{aligned}
\left(\frac{dW}{du}\right)^2
&=-\frac{K^2}{u}+\frac{K^6u}{4\alpha}\\[5pt]
&=\frac{p_+p_-}{u}+\frac{p_-^2u}{4}.
\end{aligned}
\label{eq:HJ}
\end{equation}
For the inverse metric $g^{ab}$ of \eqref{eq:reduced-metric}, this is the null Hamilton--Jacobi equation:
\begin{equation}
g^{ab}\partial_a\mathcal S\,\partial_b\mathcal S
=4u^2
\left[\left(\frac{dW}{du}\right)^2
-\frac{p_+p_-}{u}-\frac{p_-^2u}{4}\right]=0.
\label{eq:null-condition-S}
\end{equation}

The WKB phase $\mathcal S$ is therefore Hamilton's principal function for null propagation, and the corresponding null geodesics are its characteristics. Their tangent is proportional to $g^{ab}\partial_b\mathcal S$, with a constant proportionality factor fixing the affine normalization. Under the barrier, $0<u<u_*$, the decaying solution has $dW/du=i|dW/du|$. Choosing the factor $-i$ makes $u$ increase with a real affine parameter:
\begin{equation}
g_{ab}\dot x^b=-i\,\partial_a\mathcal S.
\label{eq:geodesic-from-phase}
\end{equation}
Here the dot denotes differentiation with respect to that affine parameter. The conserved geodesic momenta are consequently $-ip_\pm$, as in \eqref{eq:imaginary-momenta}, while the Fourier momenta $p_\pm$ in \eqref{eq:eikonal-phase} remain real. The radial component of \eqref{eq:geodesic-from-phase} is
\begin{equation}
\frac{\dot u}{4u^2}
=-i\frac{dW}{du}=\left|\frac{dW}{du}\right|>0.
\label{eq:geodesic-from-phase-components}
\end{equation}
Squaring \eqref{eq:geodesic-from-phase-components} and using \eqref{eq:HJ} gives \eqref{eq:complex-geodesic}. Together with \eqref{eq:imaginary-momenta}, these are exactly the equations of the geodesic constructed in Section~\ref{sec:reduced-metric}. 

The turning point \eqref{eq:geodesic-turning-point} satisfies $u_*=2\sqrt\alpha/K^2$, or $s=1$ by \eqref{eq:barrier-coordinate}. On the returning segment, the root $dW/du=-i|dW/du|$ of \eqref{eq:HJ} gives $\dot u<0$ in \eqref{eq:geodesic-from-phase}.

Contracting \eqref{eq:geodesic-from-phase} with $\dot x^a$ gives $\dot x^a\partial_a\mathcal S=i g_{ab}\dot x^a\dot x^b=0$. Hence $\mathcal S$ is constant along the geodesic:
\begin{equation}
p_+\,dx^++p_-\,dx^-=-\frac{dW}{du}\,du.
\label{eq:phase-constant}
\end{equation}
We define the eikonal phase using the Fourier momenta in \eqref{eq:eikonal-phase}:
\begin{equation}
\delta=-p_+\Delta x^+-p_-\Delta x^-.
\label{eq:phase-endpoint-definition}
\end{equation}
On both segments $dW=i|dW/du|\,|du|$. Integrating \eqref{eq:phase-constant} therefore gives
\begin{equation}
\begin{aligned}
\delta
&=2i\int_0^{u_*}\left|\frac{dW}{du}\right|du\\[5pt]
&=2iI=iA\alpha^{1/4}.
\end{aligned}
\label{eq:geodesic-phase-result}
\end{equation}
The factor of two accounts for the equal contributions of the outgoing and returning segments. The last line uses \eqref{eq:barrier-coordinate}, \eqref{eq:one-crossing} and \eqref{eq:geodesic-action-value}. Substitution of the endpoint separations \eqref{eq:positive-plus} and \eqref{eq:positive-minus} into \eqref{eq:phase-endpoint-definition} gives the same result. Thus $e^{i\delta}=e^{-A\alpha^{1/4}}$ reproduces the WKB absorption factor \eqref{eq:retarded-rate}. The prefactors of $\Im f_T$ and $\Im f_L$ follow from the matching in Appendix~\ref{app:matching}.

\subsection{Geodesic endpoints and the OPE singularity}
\label{sec:complex-geodesic}

The geodesic of Section~\ref{sec:reduced-metric} is related by symmetry to the near-lightcone limit of the geodesic in Ref.~\cite{Araya:2026shz}. For any nonzero complex constant $\lambda$, the complexified metric \eqref{eq:reduced-metric} is invariant under
\begin{equation}
\begin{aligned}
x^+&\to\lambda x^+,\\[4pt]
x^-&\to\lambda^{-3}x^-,\\[4pt]
u&\to\lambda^{-2}u.
\end{aligned}
\label{eq:scaling-isometry}
\end{equation}
Setting $\lambda=i$ gives
\begin{equation}
\begin{aligned}
x^\pm&\to ix^\pm,\\[4pt]
u&\to-u.
\end{aligned}
\label{eq:scaling-i}
\end{equation}

The separations \eqref{eq:positive-plus} and \eqref{eq:positive-minus} become real:
\begin{equation}
\begin{aligned}
\Delta x^+&=\frac{3A}{4}\left(\frac{-p_-}{p_+}\right)^{1/4}>0,\\[6pt]
\Delta x^-&=-\frac{A}{4}\left(\frac{p_+}{-p_-}\right)^{3/4}<0.
\end{aligned}
\label{eq:real-endpoints}
\end{equation}
This geodesic turns at $u=-u_*$. Since $u\propto r^{-2}$, negative $u$ means imaginary $r$, where $r$ is the usual radial coordinate introduced after \eqref{eq:metric}. Its endpoints are real and spacelike separated, as for the imaginary geodesic of Ref.~\cite{Araya:2026shz} in the near-lightcone limit.

The endpoint invariant, in the normalization of that reference, is
\begin{equation}
\mathcal X=-16\,\Delta x^-(\Delta x^+)^3.
\label{eq:source-position-variable}
\end{equation}
The factor $16$ accounts for its lightcone separations being twice ours. Equation~\eqref{eq:scaling-isometry} leaves $\mathcal X$ unchanged, so either set of endpoints gives
\begin{equation}
\begin{aligned}
\mathcal X&=\frac{27A^4}{16}\\[5pt]
&=\frac{4\pi^2}{3}
\left[\frac{\Gamma(1/4)}{\Gamma(3/4)}\right]^4
=\mathcal X_c.
\end{aligned}
\label{eq:calX}
\end{equation}
Here $\mathcal X_c$ is the known singularity location of the scalar stress-tensor sector~\cite{Araya:2026shz}.

\section{Discussion}
\label{sec:discussion}

In this paper we consider  various aspects of the thermal current correlator in the near-lightcone limit.
This is the DIS regime in which the virtuality and the plasma compete, 
 and  the dependence on the current--stress-tensor coupling reduces to the factors $r_T$ and $r_L$.
Exponentially weak absorption is related to propagation along a complex null geodesic. 

\subsection{The plasma in the near-lightcone limit}
\label{sec:disc-geometry}

Section~\ref{sec:kaigorodov} gives a geometric description of the near-lightcone limit. The black-brane metric becomes the Kaigorodov metric \eqref{eq:kaigorodov}, which can also be obtained by taking an infinite boost and sending the temperature to zero while keeping the boosted null momentum density fixed~\cite{Cvetic:1998jf,Brecher:2000pa,Singh:2010zs,Armas:2026mul}. In the limiting state only $\langle T_{++}\rangle$ is nonzero. The fast probe therefore sees a uniform null energy density. The fixed-$\alpha$ limit of the thermal correlator describes its propagation through this state.

Eq.  \eqref{eq:wave-vertex} explains the simple dependence on the Weyl coupling. The thermal term multiplies $F_{-\perp}^2$ in the transverse channel and $F_{-\zeta}^2$ in the longitudinal channel. The Weyl interaction changes their coefficients to $r_T$ and $r_L$, while leaving their radial dependence unchanged. 
This gives rise to  the argument rescaling \eqref{eq:response-universality}. 
At conformal collider bounds one coefficient vanishes, and the corresponding channel has its vacuum form.\footnote{Going close to the collider bound in the holographic model is incompatible with a large gap to higher-spin operators~\cite{Camanho:2014apa,Kulaxizi:2017ixa,Costa:2017twz,Meltzer:2017rtf}.}

\subsection{Universality and the thermal OPE}
\label{sec:disc-universal}

The rescaling of the arguments in $f_T$ and $f_L$ has a simple consequence for their large-$\alpha$ expansions: the coefficient of $\alpha^{-n}$  gets multiplied by $r_{T,L}^{\,n}$. 
This translates into rescaling of the leading-twist multi-stress tensor thermal coefficients.
Denote by $[T^n]_{2n}$ the operators made from $n$ stress tensors with spin $2n$ and dimension $4n$ at leading order in large $N$. These are the multi-stress tensors that survive the near-lightcone limit~\cite{Fitzpatrick:2019zqz,Esper:2023jeq}.

A field-theory derivation of the universality would be useful. 
The stress-tensor contribution to the current two-point function is fixed by conformal symmetry and the Ward identity. 
The two polarization factors are $1-\hat a_2/3$ and $1+2\hat a_2/3$ in any CFT~\cite{Hofman:2008ar}. 
For scalar external operators, crossing has been used to determine leading-twist multi-stress-tensor couplings under holographic spectrum assumptions~\cite{Karlsson:2019qfi,Kulaxizi:2019tkd,Karlsson:2019dbd,Li:2019zba} and implies universality of the leading-twist thermal coefficients.
It would be interesting to repeat these calculations for external currents.
 
\subsection{Moments of the structure functions}
\label{sec:disc-moments}

The coefficients of the thermal expansion can also be obtained from absorption. To state this relation, set $\gamma=0$ and define $c_n^{(T)}$ and $c_n^{(L)}$ by the asymptotic expansions, valid as $|\alpha|\to\infty$ in any closed sector excluding the positive real axis (Appendix~\ref{app:moments}),
\begin{equation}
\begin{aligned}
f_T(\alpha)-f_T(\infty)&\sim\sum_{n\ge1}\frac{c_n^{(T)}}{\alpha^n},\\
f_L(\alpha)-f_L(\infty)&\sim\sum_{n\ge1}\frac{c_n^{(L)}}{\alpha^n}.
\end{aligned}
\label{eq:disc-cn-def}
\end{equation}
The constants $f_T(\infty)=2\gamma_E-1$ and $f_L(\infty)=-2\gamma_E$ are the vacuum values in \eqref{eq:current-endpoint-responses}, with $\gamma_E$ Euler's constant. The first coefficients are $c_1^{(T)}=1/20$ and $c_1^{(L)}=1/30$, as derived in Appendix~\ref{app:moments}.

Appendix~\ref{app:moments} derives a dispersion relation in $\alpha$ and obtains
\begin{equation}
c_n^{(i)}=\frac1\pi\int_0^\infty\dd\alpha\,\alpha^{n-1}\Im f_i(\alpha),
\qquad i=T,L.
\label{eq:disc-moments}
\end{equation}
Thus each coefficient in the real asymptotic expansion is a moment of $\Im f_i$. The signs follow from the conserved radial flux. Appendix~\ref{app:matching} shows that the incoming solution has $\Im f_T>0$ and $\Im f_L>0$. Equation~\eqref{eq:disc-moments} then makes every $c_n^{(T)}$ and $c_n^{(L)}$ positive.

The same relation can be written in DIS variables. At fixed $Q$ and $T$, \eqref{eq:dict2} gives $x=4\pi^2T\sqrt\alpha/Q$, and \eqref{eq:current-structure} relates $\Im f_i$ to $F_T$ and $F_L$. Consequently $c_n^{(i)}$ is proportional to the moment $\int_0^\infty\dd x\,x^{2n-2}F_i(x)$; the normalization is given in Appendix~\ref{app:moments}. The $n=1$ relations reproduce the sum rules of Ref.~\cite{Hatta:2007cs}.
At nonzero Weyl coupling the $n$-th moment is multiplied by $r_i^n$.

Spectral positivity  holds for a thermal state in any unitary theory. 
If a theory also admits the dispersion relation \eqref{eq:disc-moments}, this positivity constrains its thermal coefficients. 
In the setup of this paper this also implies $r_T\ge0$ and $r_L\ge0$. 
This is the thermal analogue of the relation between DIS positivity and conformal collider bounds in Ref.~\cite{Komargodski:2016gci}, which considers a hadronic target in a gapped phase.

\subsection{Open questions}
\label{sec:disc-open}

\paragraph{Resummation and the absorptive prefactor.}
The series \eqref{eq:disc-cn-def} has real coefficients, while the retarded functions have an exponentially small imaginary part. The incoming and outgoing solutions share the asymptotic series \eqref{eq:disc-cn-def} and differ at large positive $\alpha$ only by this exponentially small contribution. The factorial growth in \eqref{eq:large-order-series} leads to a Borel singularity at $A$ when the expansion is organized in the small WKB parameter $\alpha^{-1/4}$~\cite{Mueller:2008bt}. Because the singularity lies on the integration contour, a resummation must specify how the contour passes it. WKB matching determines the leading absorptive prefactor. An exact WKB analysis of the reduced equations could relate the resummation choice directly to the incoming prescription and determine the systematic subleading corrections to that prefactor. Recent work on thermal correlators provides the relevant methods~\cite{Afkhami-Jeddi:2025wra,Giombi:2026kdz,Jia:2025jbi,Jia:2026ryl,Grassi:2026ekz}.

Equation~\eqref{eq:calX} relates the coefficient $A$ in the absorption exponent to the position-space singularity location $\mathcal X_c$ through $A^4=16\mathcal X_c/27$. A further question is whether the power and normalization of the stress-tensor-sector singularity can also be related to the prefactor multiplying $e^{-A\alpha^{1/4}}$, which can be computed independently by WKB matching. For the current, this requires determining the detailed singularity of its stress-tensor sector, as discussed in Section~\ref{sec:disc-universal}. It gives a concrete setting for the question raised in Ref.~\cite{Grassi:2026ekz} about the relation between position-space singularities and the Borel singularities of momentum-space thermal correlators.

\paragraph{The lightlike limit and photon emission.}
At the opposite end of the DIS crossover, $\alpha\to0$, absorption is strong. The transverse coefficient behaves as $\Im f_T\propto\alpha^{-1/3}$ by \eqref{eq:current-small-alpha-absorption}. After restoring the kinematic factor in \eqref{eq:current-structure}, the transverse spectral function has a finite lightlike limit, independent of the virtuality, and grows as $T^{4/3}k^{2/3}$ at large $k/T$. At $\omega=k$ this is the quantity that determines the emission rate of photons coupled to the current~\cite{Caron-Huot:2006pee}.

The argument rescaling in \eqref{eq:response-universality} predicts a factor $r_T^{1/3}$. More explicitly, with $R_1$ the transverse current correlator of Section~\ref{sec:dictionary},
\begin{equation}
\frac{\Im R_1(\omega=k,k;\gamma)}
     {\Im R_1(\omega=k,k;0)}
\longrightarrow r_T^{1/3}
=\left(1-\frac{\hat a_2}{3}\right)^{1/3},
\qquad k/T\to\infty,
\label{eq:disc-photon-factor}
\end{equation}
for fixed $r_T>0$ and fixed current normalization.

\paragraph{Timelike momentum.}
For $\omega>k$, the vacuum current correlator already has a nonzero imaginary part. The quantity of interest is then the thermal correction to this vacuum contribution. We consider the transverse channel at $\gamma=0$, first taking the near-lightcone limit and then large $|\alpha|$.

The vacuum value follows directly from the boundary logarithm. With $C_T$ the source in \eqref{eq:current-boundary} and $v=K^2u$, the vacuum solution has the expansion
\begin{equation}
\frac{A_\perp^{\rm vac}(v)}{C_T}
=1+v\big[\log v+2\gamma_E-1\big]+O(v^2\log v).
\label{eq:disc-timelike-vacuum-profile}
\end{equation}
For timelike momentum, $K^2<0$, so $v=K^2u<0$ along the physical radial interval $u>0$. The retarded prescription $\omega\to\omega+i0$ takes $Q^2=k^2-(\omega+i0)^2$, and hence $v$, to the negative real axis from below. Define $f_T^{\rm tl}$ by writing the boundary expansion \eqref{eq:current-boundary} with $\log|v|$; the superscript specifies the physical timelike continuation. Thus
\begin{equation}
\log v=\log|v|-i\pi,\qquad
f_{T,\rm vac}^{\rm tl}=2\gamma_E-1-i\pi.
\label{eq:disc-timelike-vacuum}
\end{equation}
This is the origin of the vacuum value $\Im f_T^{\rm tl}=-\pi$. We solve the transverse equation in \eqref{eq:reduced} at timelike momentum with the incoming boundary condition, and extract $f_T^{\rm tl}$ from the boundary expansion. The numerical results are consistent with the large-$|\alpha|$ behavior
\begin{equation}
\Im f_T^{\rm tl}
 =-\pi-2\pi e^{-(A/\sqrt2)|\alpha|^{1/4}}
\left[\cos\!\left(\frac{A}{\sqrt2}|\alpha|^{1/4}\right)
+O\!\left(|\alpha|^{-1/4}\right)\right].
\label{eq:disc-timelike}
\end{equation}
The power corrections in $1/\alpha$ are real. The imaginary part therefore oscillates about its vacuum value as the momentum varies, with an amplitude that decays exponentially in $|\alpha|^{1/4}$.

Exponentially damped oscillations also appear in the timelike thermal scalar correlators studied in Refs.~\cite{Jia:2025jbi,Jia:2026ryl}. Let $G_R$ denote the scalar retarded correlator and $\delta\Im G_R$ its leading exponentially small contribution to the imaginary part. Set $w=\omega/(2\pi T)$ and $\eta=k/\omega$, with $0<\eta<1$. We denote by $\mathcal V(\eta)$ the function $v$ in eq.~(5.8) of Ref.~\cite{Jia:2026ryl}. At large $w$ and fixed $\eta$, eqs.~(6.5)--(6.7) of that reference give, up to a power-law prefactor,
\begin{equation}
\delta\Im G_R
\propto e^{-w[\pi-\mathcal V(\eta)]}
\cos\!\bigl(w[\pi-\mathcal V(\eta)]\bigr).
\label{eq:disc-scalar-oscillations}
\end{equation}
The exponential determines the suppression of the correction, while the cosine describes its oscillations as the momentum varies. The same quantity $w[\pi-\mathcal V(\eta)]$ appears in both factors. Rewriting eq.~(5.8) of Ref.~\cite{Jia:2026ryl} in integral form gives
\begin{equation}
\pi-\mathcal V(\eta)
=\frac{3A}{8}\int_{\eta^2}^{1}\frac{dy}{[y(1-y)]^{1/4}}
\simeq\frac{A}{2}(1-\eta^2)^{3/4},
\qquad \eta\to1.
\label{eq:disc-JR-endpoint}
\end{equation}
Equation~\eqref{eq:dict1} gives $|\alpha|=w^4(1-\eta)^3(1+\eta)$. Multiplying \eqref{eq:disc-JR-endpoint} by $w$ therefore yields
\begin{equation}
\begin{aligned}
w\big[\pi-\mathcal V(\eta)\big]
&\simeq\frac{A}{2}\sqrt{1+\eta}\,|\alpha|^{1/4}\\
&\longrightarrow\frac{A}{\sqrt2}|\alpha|^{1/4},
\qquad \eta\to1,\quad |\alpha|\ \text{fixed}.
\end{aligned}
\label{eq:disc-JR-timelike-scale}
\end{equation}
{\emergencystretch=3em
Substituting \eqref{eq:disc-JR-timelike-scale} into \eqref{eq:disc-scalar-oscillations} gives the exponential $e^{-(A/\sqrt2)|\alpha|^{1/4}}$ and the cosine $\cos[(A/\sqrt2)|\alpha|^{1/4}]$ that appear in the current result \eqref{eq:disc-timelike}. The same coefficient $A$ therefore determines the spacelike exponent \eqref{eq:retarded-rate} and the timelike decay and oscillation scales. At leading WKB order, both the scalar and gauge-field equations reduce to the null Hamilton--Jacobi equation, which explains the agreement of these exponential scales. 
\par}

\paragraph{Euclidean correlators and lattice QCD.}
Euclidean current correlators offer a way to compare these scaling predictions with QCD. At a Matsubara frequency $\omega=i\omega_n$, with $\omega_n=2\pi nT$ and $n$ a positive integer, keeping the spacelike virtuality $Q^2$ fixed requires $k^2=Q^2-\omega_n^2$. When $\omega_n>Q$, the spatial momentum must therefore be imaginary~\cite{Meyer:2018xpt,Ce:2020wgg}:
\begin{equation}
\omega=i\omega_n,\qquad k=ik_E,\qquad
k_E=\sqrt{\omega_n^2-Q^2},\qquad 0<k_E<\omega_n.
\label{eq:disc-euclidean-momenta}
\end{equation}
Here $k_E$ is real. If $\omega_n<Q$, fixed virtuality instead requires real spatial momentum. The regime $\omega_n\gg Q$ uses $k_E\simeq\omega_n$: the two large contributions to $Q^2$ nearly cancel, as in the Lorentzian near-lightcone limit.

Imaginary momentum is implemented by weighting the Euclidean correlator at real spatial separation $z$ by $e^{-ikz}=e^{k_Ez}$ in its Fourier integral. Here $z$ is the boundary spatial coordinate along the momentum. This choice allows the Euclidean dispersion relation to involve DIS structure functions at the same fixed virtuality $Q^2$ \cite{Ce:2020wgg}. The continued virtuality and near-lightcone parameter are
\begin{equation}
Q^2=\omega_n^2-k_E^2>0,\qquad
\alpha=-\frac{Q^6}{(2\pi T)^4(\omega_n+k_E)^2}<0.
\label{eq:disc-euclidean-alpha}
\end{equation}
For $\alpha<0$, regularity at the Euclidean horizon selects bulk solutions $A_\perp(v)$ and $a(v)$ that decay at large $v$; both solutions are real for real source coefficients $C_T$ and $C_L$.

To test the holographic prediction, one would subtract the vacuum contribution from the Euclidean current correlators and divide by the known tensor and kinematic prefactors. At fixed temperature $T$, the resulting $f_T-f_T(\infty)$ and $f_L-f_L(\infty)$ should approach fixed values as $Q/T$ increases at fixed $\alpha<0$, defined by \eqref{eq:disc-euclidean-alpha}. Comparing lattice data at different Matsubara frequencies and virtualities but the same $\alpha$ would test this prediction.

\acknowledgments
I thank Edmond Iancu, Manuela Kulaxizi and Mukund Rangamani for comments on the draft.
I thank Queen Mary University of London for hospitality. This work was supported in part by Taighde \'Eireann -- Research Ireland under Grants SFI-22/FFP-P/11444 and 22/EPSRC/3832. I acknowledge the use of ChatGPT and Claude.

\appendix
\section{Equation in the longitudinal channel}
\label{app:reduced}

The longitudinal perturbations involve both $A_t$ and $A_z$. They are related by a constraint, so they do not describe two independent response functions. This appendix first uses that constraint to obtain the black-brane equation for $a=\dd A_t/\dd u$, then derives the longitudinal equation directly in the limiting geometry and matches the two variables.

\subsection{The black-brane equation}

Let $\nabla_M$ denote the covariant derivative associated with the bulk metric. Varying the gauge-field action \eqref{eq:weyl-action} gives
\begin{equation}
\nabla_M\left(F^{MN}-4\gamma L^2C^{MNPQ}F_{PQ}\right)=0.
\label{eq:weyl-eom}
\end{equation}
One can use the Fourier convention and radial gauge of Section~\ref{sec:thermal-DIS}, and the functions $h_T,h_L$ defined in \eqref{eq:weyl-factors}. The radial component contains a constraint between the two fields. The temporal component supplies the other relation we need:
\begin{align}
w h_L a+qfh_T\frac{\dd A_z}{\dd u}&=0,
\label{eq:weyl-constraint}\\
h_T(q^2A_t+qwA_z)&=uf\frac{\dd(h_La)}{\dd u}.
\label{eq:weyl-temporal}
\end{align}
These are the corresponding component equations of~\cite{Ritz:2008kh}, whose dimensionful Weyl coupling is $\gamma L^2$ in our notation. The spatial component of \eqref{eq:weyl-eom} along $z$ follows from these two because the divergence of \eqref{eq:weyl-eom} vanishes identically, so \eqref{eq:weyl-constraint} and \eqref{eq:weyl-temporal} contain the complete content of the equations of motion in this channel.

Define
\[
\Phi=q^2A_t+qwA_z.
\]
We will express both $\Phi$ and its derivative in terms of $a$. The temporal equation gives the first expression directly:
\begin{equation}
\Phi=\frac{uf}{h_T}\frac{\dd(h_La)}{\dd u}.
\end{equation}
For the derivative, first differentiate the definition of $\Phi$ and then use the radial constraint to eliminate $\dd A_z/\dd u$:
\begin{equation}
\frac{\dd\Phi}{\dd u}
=q^2a+qw\frac{\dd A_z}{\dd u}
=\left(q^2-\frac{w^2h_L}{fh_T}\right)a.
\end{equation}
Finally, differentiate the expression for $\Phi$ obtained from the temporal equation and equate it to the expression for $\dd\Phi/\dd u$ above. This yields \eqref{eq:weyl-full-L}, a second-order equation containing $a$ alone. At the boundary, where $f$, $h_T$ and $h_L$ tend to one, \eqref{eq:weyl-temporal} gives $q^2A_t(0)+qwA_z(0)=\lim_{u\to0}u\,\dd a/\dd u$, which is the source coefficient $C_L$ of \eqref{eq:longitudinal-source}.

Using $a$ has a practical advantage. The equation for the gauge-invariant longitudinal field $E_z=wA_z+qA_t$ has a first-derivative coefficient that diverges where $w^2h_L=q^2fh_T$, the local longitudinal light cone; this is an apparent singularity of the equation, not of its solutions. The equation for $a$ has singular points only at $u=0$, $u=1$ and where $h_L=0$. For the allowed range of $\hat a_2$ the last point lies outside the bulk, since $h_L$ vanishes at $u^2=1/(24\gamma)\ge4/3$ for $\gamma>0$ and nowhere for $\gamma<0$. This makes the boundary expansion and the numerical integration simpler.

As a check, set the Weyl coupling to zero. The functions $h_T$ and $h_L$ then equal one, and the equation becomes
\begin{equation}
\frac{\dd^2a}{\dd u^2}+\frac{(uf)'}{uf}\frac{\dd a}{\dd u}
+\frac{w^2-q^2f}{uf^2}a=0,
\end{equation}
where the prime denotes a $u$ derivative. This is the longitudinal equation at $\gamma=0$ used in the thermal DIS calculation~\cite{Hatta:2007cs}.

\subsection{The longitudinal variable in the Kaigorodov geometry}

Let us use the limiting coordinates of Section~\ref{sec:kaigorodov}, radial gauge $A_u=0$, and consider a Fourier mode $e^{i(p_+x^++p_-x^-)}$. 
In this subsection a prime denotes $d/du$, with $u=\zeta^2$. 
Define
\begin{equation}
\psi=A_-',\qquad E=p_+A_--p_-A_+.
\label{eq:kaigorodov-longitudinal-fields}
\end{equation}
Here $A_\pm$ are the covariant lightcone components of the gauge potential, and $E$ is invariant under gauge transformations.
 The radial constraint and the $N=+$ component of the gauge-field equation give, respectively,
\begin{align}
2p_+A_-'+2p_-A_+'+r_Lp_-u^2A_-'&=0,
\label{eq:kaigorodov-constraint}\\
u\psi'&=-\frac{p_-}{2}E.
\label{eq:kaigorodov-plus-equation}
\end{align}
The constraint eliminates $A_+'$ and therefore determines
\begin{equation}
E'=\left(2p_++\frac{r_Lp_-u^2}{2}\right)\psi.
\end{equation}
Differentiating \eqref{eq:kaigorodov-plus-equation} now yields a single equation for $\psi$:
\begin{equation}
\psi''+\frac{\psi'}u
=\left[-\frac{p_+p_-}{u}-\frac{r_Lp_-^2u}{4}\right]\psi.
\label{eq:kaigorodov-longitudinal-u}
\end{equation}
In the coordinate $\zeta$ this is \eqref{eq:longitudinal-kaigorodov}. The source coefficient is $C_\psi=\lim_{u\to0}u\psi'=-(p_-/2)E(0)$; the factor relating it to $E(0)$ is independent of $\gamma$.

To match $\psi$ to the earlier field $a$, distinguish the original black-brane coordinate $U$ and dimensionless lightcone coordinates $\bar x^\pm$ from the limiting coordinates $u,x^\pm$. Let $\bar A_\pm$ and $\bar p_\pm$ be the original lightcone gauge components and momenta. The coordinate and momentum transformations are
\begin{equation}
\begin{gathered}
U=\epsilon^2u,\qquad
\bar x^+=\epsilon^{-1}x^+,\qquad
\bar x^-=\epsilon^3x^-,\\
\bar p_+=\epsilon p_+,\qquad
\bar p_-=\epsilon^{-3}p_-.
\end{gathered}
\label{eq:kaigorodov-coordinate-limit}
\end{equation}
The gauge potential is a one-form, so $\bar A_+=\epsilon A_+$ and $\bar A_-=\epsilon^{-3}A_-$. Since the original physical temporal component is $A_t=\pi T(\bar A_++\bar A_-)$, the field $a=dA_t/dU$ satisfies
\begin{equation}
\frac{\epsilon^5}{\pi T}a=\psi+\epsilon^4A_+'.
\label{eq:kaigorodov-a-matching}
\end{equation}
At fixed $u,p_\pm$ the constraint \eqref{eq:kaigorodov-constraint} makes the second term vanish relative to $\psi$ as $\epsilon\to0$. Thus $\psi$ is the normalized limit of $a$. The source coefficient and the finite boundary coefficient acquire the same constant normalization, so their ratio is unchanged. By \eqref{eq:current-boundary}, $f_L$ is minus this ratio and is therefore unchanged as well. In the parametrization \eqref{eq:epsilon-limit}, $p_+=1$ and $p_-=-\alpha$. More generally $\tilde u=u/p_+^2$, and both $\alpha=-p_-p_+^3$ and $v=-p_+p_-u$ are invariant under \eqref{eq:kaigorodov-coordinate-limit}.

\section{Numerical calculations at vanishing Weyl coupling}
\label{app:numerics}

We obtain $f_T^{(0)}$ and $f_L^{(0)}$ by solving \eqref{eq:universal-T} and \eqref{eq:universal-L} at $\gamma=0$. At large positive $v$, the incoming solution has the phase $\exp[+iv^{3/2}/(3\sqrt\alpha)]$. Its asymptotic expansion supplies the initial data. We integrate toward the boundary and match to two independent boundary series, including the logarithms in \eqref{eq:current-boundary}; the ratios of their coefficients determine the response functions.

To resolve exponentially small imaginary parts, we compute the conserved radial flux and divide by the squared boundary source. The transverse and longitudinal fluxes divided by $|C_T|^2$ and $|C_L|^2$ give $\Im f_T^{(0)}$ and $\Im f_L^{(0)}$, respectively, as derived in Appendix~\ref{app:matching}. This avoids extracting a tiny imaginary part by subtracting nearly equal complex numbers.

An independent calculation integrates along $v=e^{i\pi/3}y$, with $y>0$ and real $\alpha>0$. This ray lies inside $0<\arg v<2\pi/3$, where the incoming exponential is subdominant; on the ray it is $\exp[-y^{3/2}/(3\sqrt\alpha)]$. Continuing the decaying solution in this sector to the positive real $v$-axis gives the incoming solution. The reduced equations have no finite singularity away from $v=0$, so the continuation between the ray and the positive real axis crosses no singularity. We use $\log v=\log y+i\pi/3$ on the ray and continue the same boundary series to the real axis. At eight tested values spanning $10^{-3}\leq\alpha\leq10^4$, the two methods agree to within $2\times10^{-12}$ in the real parts and $2\times10^{-11}$ relatively in the nonzero imaginary parts. Changes of precision and radial cutoffs confirm convergence.

The numerical data cover $10^{-4}\leq\alpha\lesssim2.61\times10^4$. The plots interpolate the real parts and the logarithms of the positive quantities $\Im f_T^{(0)}$ and $\Im f_L^{(0)}$ as functions of $\log\alpha$. Additional solutions between grid points check the interpolation. All nonzero plotted values lie within this range; the collider endpoints use the exact vacuum values \eqref{eq:current-endpoint-responses}.

\section{Incoming-wave matching and absorption}
\label{app:matching}

This appendix explains how the incoming boundary condition gives the exponentially small imaginary parts $\Im f_T$ and $\Im f_L$. We first take the fixed-$\alpha$ near-lightcone limit at vanishing Weyl coupling, $\gamma=0$, and then study the reduced equations \eqref{eq:reduced} at large positive $\alpha$. We connect the incoming wave beyond the barrier to the boundary expansions \eqref{eq:current-boundary}, whose coefficients determine the response functions.

In the radial variable $v=K^2u$ of \eqref{eq:fixed-v}, the two Schr\"odinger fields are $\Phi_T=A_\perp$ and $\Phi_L=\sqrt v\,a$. Their equations are $d^2\Phi_i/dv^2=\widehat V_i\Phi_i$, with $i=T,L$. The potentials \eqref{eq:reduced-potentials-v} become
\begin{equation}
\widehat V_T(v)=\frac1v-\frac{v}{4\alpha},\qquad
\widehat V_L(v)=\frac1v-\frac{v}{4\alpha}-\frac1{4v^2}.
\label{eq:matching-potentials}
\end{equation}
We first consider the transverse field $A_\perp$.

\paragraph{The incoming wave and the turning point.}
The transverse potential vanishes at $v_*=2\sqrt\alpha$. For $v>v_*$, $\widehat V_T<0$ and the transverse solutions $\Phi_T(v)$ are oscillatory. The retarded boundary condition selects propagation toward increasing $v$. Define $k_v(v)=\sqrt{-\widehat V_T(v)}$ in this region and let $C$ be the normalization of the incoming wave. Away from the turning point, its WKB form is
\begin{equation}
\Phi_T(v)\simeq\frac{C}{\sqrt{k_v(v)}}
\exp\!\left[i\left(\int_{v_*}^v k_v(v')\,dv'+\frac\pi4\right)\right].
\label{eq:incoming-wkb}
\end{equation}
Near $v_*$, the potential is approximately linear, $\widehat V_T(v)\simeq(v_*-v)/(2\alpha)$. With the variable $\zeta$ defined below, the equation for $\Phi_T$ reduces to the Airy equation
\begin{equation}
\zeta=(2\alpha)^{-1/3}(v_*-v),\qquad
\frac{d^2\Phi_T}{d\zeta^2}=\zeta\Phi_T.
\label{eq:matching-Airy-equation}
\end{equation}
On the oscillatory side, where $-\zeta\gg1$, the Airy asymptotics are
\begin{equation}
\begin{aligned}
\operatorname{Bi}(\zeta)&\simeq
\frac{\cos[2(-\zeta)^{3/2}/3+\pi/4]}{\sqrt\pi\,(-\zeta)^{1/4}},\\
\operatorname{Ai}(\zeta)&\simeq
\frac{\sin[2(-\zeta)^{3/2}/3+\pi/4]}{\sqrt\pi\,(-\zeta)^{1/4}}.
\end{aligned}
\label{eq:matching-Airy-oscillatory}
\end{equation}
In the linear-potential region, $\int_{v_*}^v k_v(v')\,dv'\simeq2(-\zeta)^{3/2}/3$, so $\operatorname{Bi}+i\operatorname{Ai}$ reproduces the incoming exponential in \eqref{eq:incoming-wkb}. The common $\pi/4$ comes from \eqref{eq:matching-Airy-oscillatory} and can be absorbed into the overall normalization $C$.

Under the barrier, $\zeta>0$, the large-$\zeta$ asymptotics are
\begin{equation}
\operatorname{Ai}(\zeta)\simeq
\frac{e^{-2\zeta^{3/2}/3}}{2\sqrt\pi\,\zeta^{1/4}},\qquad
\operatorname{Bi}(\zeta)\simeq
\frac{e^{2\zeta^{3/2}/3}}{\sqrt\pi\,\zeta^{1/4}}.
\label{eq:matching-Airy-asymptotics}
\end{equation}
Thus the incoming condition fixes the relative coefficient of the decaying exponential to be $i/2$. To continue farther into the barrier, define
\begin{equation}
\kappa_v(v)=\sqrt{\widehat V_T(v)},\qquad
J(v)=\int_v^{v_*}\kappa_v(v')\,dv'.
\label{eq:matching-barrier-integral}
\end{equation}
The local Airy exponent $2\zeta^{3/2}/3$ agrees with $J(v)$ where the linear approximation to the potential is valid. Matching there gives the WKB solution
\begin{equation}
\Phi_T(v)\simeq\frac{C}{\sqrt{\kappa_v(v)}}
\left[e^{J(v)}+\frac{i}{2}e^{-J(v)}\right].
\label{eq:airy-connection}
\end{equation}
The two exponentials in \eqref{eq:airy-connection}, including their relative complex coefficient, describe the continuation of the incoming wave through the turning point.

\paragraph{Matching to the transverse boundary solution.}
The WKB approximation fails near $v=0$. We therefore match \eqref{eq:airy-connection} to a vacuum solution at an intermediate radius $v_m$. For $v_m^2/\alpha\ll1$, expanding \eqref{eq:matching-barrier-integral} gives the accumulated thermal correction
\begin{equation}
\begin{aligned}
\int_0^{v_m}\!\left[\kappa_v(v)-\frac1{\sqrt v}\right]dv
&=\int_0^{v_m}\!\left[-\frac{v^{3/2}}{8\alpha}
+O\!\left(\frac{v^{7/2}}{\alpha^2}\right)\right]dv\\
&=-\frac{v_m^{5/2}}{20\alpha}
+O\!\left(\frac{v_m^{9/2}}{\alpha^2}\right).
\end{aligned}
\label{eq:matching-integrated-correction}
\end{equation}
The local WKB condition at $v_m$ is
\begin{equation}
\frac{|\kappa_v'(v_m)|}{\kappa_v(v_m)^2}=O(v_m^{-1/2})\ll1.
\label{eq:matching-WKB-condition}
\end{equation}
Combining \eqref{eq:matching-WKB-condition} with a small integrated correction in \eqref{eq:matching-integrated-correction} gives
\begin{equation}
1\ll v_m\ll\alpha^{2/5}.
\label{eq:matching-window}
\end{equation}
This overlap also satisfies $v_m\ll v_*$ and $v_m^2/\alpha\ll1$ at large $\alpha$.

Let $I=\int_0^{v_*}\kappa_v(v')\,dv'$ be the barrier integral in \eqref{eq:one-crossing}. In the overlap region \eqref{eq:matching-window},
\begin{equation}
J(v)=I-2\sqrt v+O(v^{5/2}/\alpha).
\label{eq:matching-overlap-action}
\end{equation}
Using \eqref{eq:matching-overlap-action} and $\kappa_v(v)^{-1/2}\simeq v^{1/4}$ in \eqref{eq:airy-connection} gives
\begin{equation}
A_\perp(v)\simeq Ce^I v^{1/4}
\left[e^{-2\sqrt v}+\frac{i}{2}e^{-2I}e^{2\sqrt v}\right].
\label{eq:double-tunnel}
\end{equation}
This formula is used in the overlap region, where $v\gg1$.

Nearer the boundary, the leading equation is the vacuum equation. Its general solution is
\begin{equation}
A_\perp^{\rm vac}(v)
=\sqrt v\left[c_{T,K}K_1(2\sqrt v)+c_{T,I}I_1(2\sqrt v)\right],
\label{eq:matching-transverse-vacuum}
\end{equation}
where $K_1$ and $I_1$ are modified Bessel functions and $c_{T,K},c_{T,I}$ are constants fixed by matching to \eqref{eq:double-tunnel}. At large $v$, the two terms in \eqref{eq:matching-transverse-vacuum} behave as $v^{1/4}e^{-2\sqrt v}$ and $v^{1/4}e^{2\sqrt v}$. At small $v$, the $K_1$ term supplies the constant boundary source $C_T$, while the $I_1$ term begins at order $v$ and contributes to $f_T$ in \eqref{eq:current-boundary}. Thus the source magnitude grows as $e^I$ at fixed incoming normalization, while the relative imaginary contribution is suppressed by $e^{-2I}$, as in Ref.~\cite{Hatta:2007cs}.

\paragraph{The longitudinal boundary solution.}
The additional term $-1/(4v^2)$ in $\widehat V_L$ is essential near the boundary. For $v\geq v_m\gg1$, its correction to the barrier integral is $O(v_m^{-1/2})$; it therefore leaves the leading action $I\propto\alpha^{1/4}$ unchanged. The longitudinal outer turning point is shifted slightly from $v_*$, but its leading Airy connection is the same.

In the same overlap region \eqref{eq:matching-window}, the vacuum solution for $a$ is
\begin{equation}
a^{\rm vac}(v)
=c_{L,K}K_0(2\sqrt v)+c_{L,I}I_0(2\sqrt v),
\label{eq:matching-longitudinal-vacuum}
\end{equation}
where $c_{L,K},c_{L,I}$ are the two matching constants. Multiplication by $\sqrt v$ gives $\Phi_L$, whose two terms have the same large-$v$ exponential behavior as the transverse terms in \eqref{eq:matching-transverse-vacuum}. Near $v=0$, the $K_0$ term supplies the logarithmic source coefficient $C_L$, while the $I_0$ term contributes to the constant coefficient containing $f_L$. The boundary expansion \eqref{eq:current-boundary} therefore gives the same leading growth $e^I$ of the source magnitude at fixed incoming normalization.

\paragraph{The imaginary parts from the radial flux.}
The conserved flux relates these source magnitudes directly to the imaginary parts of the response functions. For $i=T,L$, define
\begin{equation}
\mathcal J_i=\Im\!\left(\Phi_i^*\frac{d\Phi_i}{dv}\right),
\label{eq:radial-flux}
\end{equation}
where the star denotes complex conjugation. The potentials in \eqref{eq:matching-potentials} are real, so $d\mathcal J_i/dv=0$. For the longitudinal field,
\begin{equation}
\Phi_L^*\Phi_L'=\frac12|a|^2+v a^*a',\qquad
\mathcal J_L=v\Im(a^*a').
\label{eq:matching-longitudinal-flux}
\end{equation}
The sources are $C_T=A_\perp(0)$ and $C_L=\lim_{v\to0}va'=q^2A_L(0)$, as in \eqref{eq:longitudinal-source}. The boundary expansions \eqref{eq:current-boundary} imply
\begin{equation}
\begin{aligned}
A_\perp^*A_\perp'&=|C_T|^2\big[\log v+1+f_T+o(1)\big],\\
v a^*a'&=|C_L|^2\big[\log v-f_L^*+o(1)\big],
\qquad v\to0.
\end{aligned}
\label{eq:matching-boundary-flux-products}
\end{equation}
The logarithms are real on $v>0$. Taking the imaginary parts gives
\begin{equation}
\mathcal J_T=|C_T|^2\Im f_T,\qquad
\mathcal J_L=|C_L|^2\Im f_L.
\label{eq:flux-response}
\end{equation}
These are exact identities for the reduced equations. The ratios $\mathcal J_i/|C_i|^2$ are independent of the overall normalization of $\Phi_i$. The incoming wave carries positive flux, so \eqref{eq:flux-response} fixes the signs $\Im f_T>0$ and $\Im f_L>0$.

At fixed incoming flux, the matching above gives $|C_T|^2$ and $|C_L|^2$ proportional to $e^{2I}$ at leading exponential order. Consequently,
\begin{equation}
\Im f_T=\frac{\mathcal J_T}{|C_T|^2}\propto e^{-2I},\qquad
\Im f_L=\frac{\mathcal J_L}{|C_L|^2}\propto e^{-2I}.
\label{eq:matching-flux-suppression}
\end{equation}
Since $2I=A\alpha^{1/4}$ by \eqref{eq:geodesic-action-value}, \eqref{eq:matching-flux-suppression} reproduces the two suppression rates \eqref{eq:retarded-rate}.

\section{Dispersion relation and moments}
\label{app:moments}

The coefficients $c_n$ can be computed by solving the radial equation order by order in $1/\alpha$. They can also be obtained by integrating $\Im f$. The purpose of this appendix is to establish why the two calculations agree. The nontrivial step is the dispersion relation: we must show that its contour receives no contribution from infinity or from the origin. A radial eigenfunction expansion provides this result. The calculation of the moments then follows directly.

We work at vanishing Weyl coupling, after the fixed-$\alpha$ near-lightcone limit, and consider the transverse channel. Channel labels are omitted until the final paragraph.

\subsection{Radial equation and boundary terms}

In the coordinate $v=K^2u$ of \eqref{eq:fixed-v}, let $y=A_\perp/C_T$, where $C_T$ is the source coefficient in \eqref{eq:current-boundary}. The equation \eqref{eq:reduced} and its boundary expansion are
\begin{equation}
y''=\left(\frac1v-\frac{v}{4\alpha}\right)y,
\qquad
y=1+v\log v+f(\alpha)v+O(v^2\log v).
\label{eq:mom-reduced-v}
\end{equation}
Primes denote $v$ derivatives. For real $\alpha>0$, the incoming solution has the large-$v$ behavior
\begin{equation}
y(v)\ \propto\ v^{-1/4}
\exp\!\left[\frac{iv^{3/2}}{3\sqrt\alpha}\right].
\label{eq:mom-incoming}
\end{equation}
Continuing $\alpha$ just below the positive real axis makes this exponential decay. We first use that decaying solution; its limit on the real axis gives the physical response $f(\alpha-i0)$, as in Appendix~\ref{app:numerics}.

The vacuum solution is
\begin{equation}
U(v)=2\sqrt v\,K_1(2\sqrt v)
=1+v\log v+(2\gamma_E-1)v+O(v^2\log v).
\label{eq:mom-vacuum}
\end{equation}
It has the same source as $y$. Subtracting its boundary coefficient defines
\begin{equation}
g(\alpha)=f(\alpha)-(2\gamma_E-1),
\label{eq:mom-g}
\end{equation}
with $\gamma_E$ Euler's constant. The difference $y-U$ therefore vanishes at the boundary and starts with $g(\alpha)v$.

Multiply the equation for $y$ by $U$ and subtract the vacuum equation multiplied by $y$. This gives
\begin{equation}
\frac{\dd}{\dd v}(Uy'-U'y)=-\frac{v}{4\alpha}Uy.
\label{eq:mom-Wronskian}
\end{equation}
At infinity, $Uy'-U'y$ vanishes because both solutions decay. At the boundary, $U,y\to1$, and the logarithms in $y'$ and $U'$ cancel. Equations~\eqref{eq:mom-reduced-v} and \eqref{eq:mom-vacuum} thus give $Uy'-U'y\to f-(2\gamma_E-1)=g$. Integrating \eqref{eq:mom-Wronskian} yields
\begin{equation}
\bigl[Uy'-U'y\bigr]_0^\infty
=0-g(\alpha)
=-\frac1{4\alpha}\int_0^\infty vU(v)y(v)\,\dd v.
\label{eq:mom-integrated-Wronskian}
\end{equation}
The response is consequently determined by the overlap of $y$ with the known vacuum solution:
\begin{equation}
g(\alpha)=\frac1{4\alpha}\langle U,y\rangle,
\qquad
\langle h_1,h_2\rangle=\int_0^\infty v\,h_1(v)^*h_2(v)\,\dd v.
\label{eq:mom-readout}
\end{equation}
Here the star denotes complex conjugation; $U$ is real.

\subsection{Dispersion relation}

To evaluate the overlap in \eqref{eq:mom-readout}, use the complete set of real radial eigenfunctions with zero boundary value. Label them by $\alpha'>0$, using the same parameter as in \eqref{eq:mom-reduced-v}:
\begin{equation}
-\frac{\dd^2h_{\alpha'}}{\dd v^2}+\frac{h_{\alpha'}}v
=\frac{v}{4\alpha'}h_{\alpha'},
\qquad h_{\alpha'}(0)=0,
\qquad
\langle h_{\alpha'},h_{\alpha''}\rangle=\delta(\alpha'-\alpha'').
\label{eq:mom-modes}
\end{equation}
The variable $\alpha'$ labels the radial eigenfunctions and will be integrated over. At large $v$, the real radial eigenfunctions $h_{\alpha'}(v)$ are superpositions of incoming and outgoing waves. The self-adjoint radial problem has an integrable potential at infinity when written in Schr\"odinger form; its complete continuum expansion is the standard result used here, with convergence in the radial norm defined in \eqref{eq:mom-readout}. For further mathematical details, see Ref.~\cite{Teschl:2014}.

Subtracting the equations for $y$ and $U$ gives
\begin{equation}
\left(-\frac{\dd^2}{\dd v^2}+\frac1v-\frac{v}{4\alpha}\right)(y-U)
=\frac{v}{4\alpha}U.
\label{eq:mom-source-solution}
\end{equation}
Multiply by $h_{\alpha'}$ and integrate by parts. Since $y-U$ and $h_{\alpha'}$ both have zero boundary value, and $y-U$ decays at infinity, the boundary terms vanish. Using \eqref{eq:mom-modes} gives
\begin{equation}
\langle h_{\alpha'},y-U\rangle
=\frac{\alpha'}{\alpha-\alpha'}\langle h_{\alpha'},U\rangle.
\label{eq:mom-projection}
\end{equation}
Thus $\langle h_{\alpha'},y\rangle=\alpha\langle h_{\alpha'},U\rangle/(\alpha-\alpha')$. Expanding the overlap $\langle U,y\rangle$ in these eigenfunctions turns \eqref{eq:mom-readout} into
\begin{equation}
g(\alpha)=\frac14\int_0^\infty
\frac{\langle h_{\alpha'},U\rangle^2}{\alpha-\alpha'}\,\dd\alpha'.
\label{eq:mom-spectral-integral}
\end{equation}
The numerator is nonnegative because both $h_{\alpha'}$ and $U$ are real. Completeness also fixes its integral:
\begin{equation}
\int_0^\infty\langle h_{\alpha'},U\rangle^2\,\dd\alpha'
=\langle U,U\rangle
=\int_0^\infty vU(v)^2\,\dd v=\frac15.
\label{eq:mom-completeness}
\end{equation}
Only this total overlap is needed; the radial eigenfunctions $h_{\alpha'}$ need not be computed explicitly. Their continuous dependence on positive $\alpha'$ and the decay of $U$ make the squared overlap locally continuous there.

Equation~\eqref{eq:mom-spectral-integral} is analytic away from the positive real $\alpha$ axis. On that axis the pole at $\alpha'=\alpha$ reaches the integration interval, giving different values from above and below: this is the cut in the $\alpha$ plane. Taking the imaginary part from below gives
\begin{equation}
\Im f(\alpha-i0)=\frac\pi4\langle h_\alpha,U\rangle^2,
\qquad \alpha>0.
\label{eq:mom-density}
\end{equation}
The vacuum subtraction is real, so $\Im g=\Im f$. Substitution into \eqref{eq:mom-spectral-integral} gives the dispersion relation,
\begin{equation}
g(\alpha)=\frac1\pi\int_0^\infty
\frac{\Im f(\alpha'-i0)}{\alpha-\alpha'}\,\dd\alpha',
\qquad \alpha\notin[0,\infty),
\label{eq:mom-dispersion}
\end{equation}
and \eqref{eq:mom-completeness} fixes the integrated absorption:
\begin{equation}
\frac1\pi\int_0^\infty\Im f(\alpha'-i0)\,\dd\alpha'=\frac1{20}.
\label{eq:mom-total-weight}
\end{equation}
The discontinuity of $g$, upper boundary value minus lower, is $-2i\Im f$. The sign of \eqref{eq:mom-density} agrees with the incoming flux in \eqref{eq:flux-response}.

\begin{figure}[t]
\centering
\includegraphics[width=0.43\textwidth]{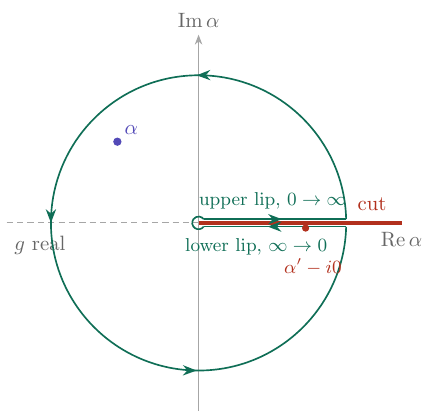}
\caption{The dispersion contour in the complex $\alpha$ plane. The retarded response is the boundary value from below the positive real cut.}
\label{fig:alpha-contour}
\end{figure}

We can now check the circle contributions in Figure~\ref{fig:alpha-contour} directly. Keep $\alpha$ fixed off the cut and take a counterclockwise circle of radius $R>|\alpha|$. Substituting \eqref{eq:mom-dispersion} gives
\begin{equation}
\frac1{2\pi i}\oint_{|z|=R}\frac{g(z)}{z-\alpha}\,\dd z
=\frac1\pi\int_0^\infty\dd\alpha'\,\Im f(\alpha'-i0)
\frac1{2\pi i}\oint_{|z|=R}\frac{\dd z}{(z-\alpha)(z-\alpha')}.
\label{eq:mom-circle-substitution}
\end{equation}
The circle ends on the two sides of the cut; the locally continuous density permits the exchange of integrals. For fixed $\alpha'<R$, both poles lie inside the circle. Their residues, $1/(\alpha-\alpha')$ and $1/(\alpha'-\alpha)$, cancel. For $\alpha'>R$, only $z=\alpha$ lies inside. Hence
\begin{equation}
\frac1{2\pi i}\oint_{|z|=R}\frac{g(z)}{z-\alpha}\,\dd z
=\frac1\pi\int_R^\infty
\frac{\Im f(\alpha'-i0)}{\alpha-\alpha'}\,\dd\alpha'.
\label{eq:mom-outer-circle}
\end{equation}
For $\alpha'>R$, $|\alpha-\alpha'|\ge R-|\alpha|$. Positivity and the finite integral \eqref{eq:mom-total-weight} therefore imply
\begin{equation}
\begin{aligned}
\left|\frac1{2\pi i}\oint_{|z|=R}\frac{g(z)}{z-\alpha}\,\dd z\right|
&\le\frac1{\pi(R-|\alpha|)}\int_R^\infty\Im f(\alpha'-i0)\,\dd\alpha'\\
&\le\frac1{20(R-|\alpha|)}\longrightarrow0.
\end{aligned}
\label{eq:mom-outer-bound}
\end{equation}
For a small counterclockwise circle of radius $r<|\alpha|$, only the poles at $z=\alpha'$ with $0<\alpha'<r$ contribute. The same reasoning gives
\begin{equation}
\left|\frac1{2\pi i}\oint_{|z|=r}\frac{g(z)}{z-\alpha}\,\dd z\right|
\le\frac1{\pi(|\alpha|-r)}\int_0^r\Im f(\alpha'-i0)\,\dd\alpha'
\longrightarrow0.
\label{eq:mom-inner-circle}
\end{equation}
The last integral tends to zero because \eqref{eq:mom-total-weight} is finite. This establishes both contour limits without using a large-$\alpha$ power series.

\subsection{Moments}

Once \eqref{eq:mom-dispersion} is established, the moments follow by expanding its denominator to any fixed finite order:
\begin{equation}
g(\alpha)\sim\sum_{n=1}^\infty\frac{c_n}{\alpha^n},
\qquad
c_n=\frac1\pi\int_0^\infty
(\alpha')^{n-1}\Im f(\alpha'-i0)\,\dd\alpha'.
\label{eq:mom-moments}
\end{equation}
Equation~\eqref{eq:mom-density} makes the integrand nonnegative, and \eqref{eq:mom-total-weight} shows that the integrated absorption is nonzero. Consequently, $c_n>0$ for every $n\ge1$.
All these integrals converge. Near $\alpha'=0$, the factor $(\alpha')^{n-1}$ is bounded for $n\ge1$, and \eqref{eq:mom-total-weight} is finite. At large $\alpha'$, the WKB suppression \eqref{eq:retarded-rate} makes $\Im f$ decay faster than any inverse power. The expansion in \eqref{eq:mom-moments} is asymptotic as $|\alpha|\to\infty$ in closed sectors away from the positive real axis.

An independent calculation solves the radial equation perturbatively. Write
\begin{equation}
y(v)\sim\sum_{n=0}^\infty\frac{y_n(v)}{\alpha^n},
\qquad y_0(v)=U(v).
\label{eq:mom-radial-expansion}
\end{equation}
Substituting into \eqref{eq:mom-reduced-v} gives
\begin{equation}
\left(-\frac{\dd^2}{\dd v^2}+\frac1v\right)y_n(v)
=\frac v4 y_{n-1}(v),\qquad n\ge1,
\label{eq:mom-recursion}
\end{equation}
with $y_n(0)=0$ and decay at infinity. The boundary identity \eqref{eq:mom-readout} then gives
\begin{equation}
c_n=\frac14\int_0^\infty vU(v)y_{n-1}(v)\,\dd v,
\qquad c_1=\frac14\langle U,U\rangle=\frac1{20}.
\label{eq:mom-radial-coefficients}
\end{equation}
This computation uses only the radial equation; no radial eigenfunctions are needed. The dispersion relation proves its equality to the absorption integral in \eqref{eq:mom-moments}.

The moments $c_n$ contain the $JJ$ OPE coefficients, thermal expectation values, and the Fourier-transform and polarization factors. Table~\ref{tab:moments} compares the radial recursion \eqref{eq:mom-recursion} with numerical integration of $\Im f$. The contribution above the tabulated $\alpha$ range is negligible for $n\le6$.

\begin{table}[t]
\centering\small
\begin{tabular}{@{}c l l@{}}
\toprule
$n$ & $c_n$, perturbative & $c_n$, moment \\
\midrule
1 & $1/20=0.05$ & $0.0500$ \\
2 & $17/280=0.0607143$ & $0.0607143$ \\
3 & $2241/2860=0.7835664$ & $0.7835664$ \\
4 & $42.45133$ & $42.45133$ \\
5 & $6562.362$ & $6562.362$ \\
6 & $2.323051\times10^{6}$ & $2.323051\times10^{6}$ \\
\bottomrule
\end{tabular}
\caption{Two independent calculations of the coefficients $c_n$. The second column is obtained by solving the radial equation perturbatively using \eqref{eq:mom-recursion} and extracting $c_n$ through \eqref{eq:mom-radial-coefficients}. The third column evaluates the moments of the numerical $\Im f$ using \eqref{eq:mom-moments}. Their agreement provides a numerical check of the dispersion relation.}
\label{tab:moments}
\smallskip
\begin{minipage}{\textwidth}
\normalsize
The first three coefficients are given as exact rational numbers; for $n\ge4$, decimal values are shown.
\end{minipage}
\end{table}

The dispersion relation \eqref{eq:mom-dispersion} has also been compared with direct solutions at negative $\alpha$. The agreement is better than one part in $10^6$ for $|\alpha|\ge1$. In this comparison the region $\alpha'<10^{-4}$ is extrapolated using \eqref{eq:current-small-alpha-absorption} and fitted subleading terms.

\subsection{DIS sum rules and large orders}

Let $F$ denote the transverse structure function. Using $x=4\pi^2T\sqrt\alpha/Q$ from \eqref{eq:dict2} and the normalization \eqref{eq:current-structure}, the moments become
\begin{equation}
\int_0^\infty\dd x\,x^{2n-2}F
=\frac{N^2Q^2}{64\pi^2}\left(\frac{4\pi^2T}{Q}\right)^{2n}c_n.
\label{eq:mom-xmoments}
\end{equation}
The spin-$2n$, dimension-$4n$ operators that survive the near-lightcone limit determine the $(2n-2)$-th moment of $F$. Their twist increases with $n$, giving the scaling $(T/Q)^{2n-2}$ at fixed $T$.

For $n=1$, \eqref{eq:mom-xmoments} and $c_1=1/20$ give
\begin{equation}
\int_0^\infty F(x)\,\dd x=\frac{\mathcal E+P}{40T^2},
\qquad \mathcal E+P=\frac{\pi^2N^2T^4}{2},
\label{eq:mom-sumrules}
\end{equation}
where $\mathcal E$ is the energy density and $P$ is the pressure. With $\mathcal E=3P$, this is the transverse combination of the sum rules (3.21)--(3.22) of Ref.~\cite{Hatta:2007cs}.

The derivation in Ref.~\cite{Hatta:2007cs} uses the single-graviton contribution and analyticity in $1/x$, with cuts assumed at $|x|<1$. Here \eqref{eq:mom-dispersion} follows from the reduced radial equation and fixes every moment. Appreciable absorption occurs at $x\lesssim T/Q$, so extending the $x$ integral beyond one adds an exponentially small contribution in the near-lightcone regime. At nonzero Weyl coupling, the rescaling \eqref{eq:response-universality} multiplies the $n$-th moment by $(1-\hat a_2/3)^n$.

At large $n$, the factor $(\alpha')^{n-1}$ in \eqref{eq:mom-moments} makes the integral sensitive to the exponentially suppressed tail of $\Im f$. Using the WKB factor $e^{-A(\alpha')^{1/4}}$ in \eqref{eq:retarded-rate}, and approximating its prefactor by a constant, gives
\begin{equation}
c_n\ \propto\ \frac{\Gamma(4n)}{A^{4n}},
\qquad
\frac{c_{n+1}}{c_n}\simeq\frac{\Gamma(4n+4)}{\Gamma(4n)A^4},
\label{eq:mom-large-n}
\end{equation}
where $A$ is given in \eqref{eq:geodesic-action-value}. A power of $\alpha'$ in the prefactor shifts the gamma-function argument; the leading factorial growth and the scale $A$ are unchanged.

Equation~\eqref{eq:mom-large-n} reproduces the growth \eqref{eq:large-order-series} found numerically in Ref.~\cite{Mueller:2008bt} and the Borel singularity \eqref{eq:borel}. The predicted ratios are $54.8$ at $n=3$ and $355.6$ at $n=5$, compared with $54.2$ and $354.0$ from Table~\ref{tab:moments}.

The longitudinal channel follows the same steps, starting from the second equation in \eqref{eq:reduced}. Its vacuum solution is $U_L=-2K_0(2\sqrt v)$, the inner product has weight $v^2$, and the homogeneous boundary condition excludes the $\log v$ term. The logarithmic source reverses the Wronskian sign, which is compensated by the minus sign multiplying $f_L$ in \eqref{eq:current-boundary}. The positive density is therefore proportional to $\Im f_L$. With $g_L=f_L+2\gamma_E$, \eqref{eq:mom-dispersion} holds with $f,g$ replaced by $f_L,g_L$. The longitudinal moments in \eqref{eq:disc-moments} are positive, with $c_1^{(L)}=1/30$. For $F_L$, \eqref{eq:mom-xmoments} holds with $c_n$ replaced by $c_n^{(L)}$. In particular, at $\gamma=0$,
\begin{equation}
\int_0^\infty F_L(x)\,\dd x=\frac{\mathcal E+P}{60T^2}.
\label{eq:mom-longitudinal-sumrule}
\end{equation}
 Restoring the Weyl coupling through \eqref{eq:response-universality}, the coefficients of $1/\alpha$ are
\begin{equation}
c_1^{(T)}=\frac{r_T}{20},\qquad
c_1^{(L)}=\frac{r_L}{30}.
\label{eq:mom-collider}
\end{equation}
Nonnegative absorption, $\Im f_T\ge0$ and $\Im f_L\ge0$, together with the dispersion relation \eqref{eq:mom-dispersion}, requires $r_T\ge0$ and $r_L\ge0$. With the definitions \eqref{eq:thermal-factor-T} and \eqref{eq:thermal-factor-L}, these are precisely the conformal collider bounds $-3/2\le \hat a_2\le3$ in \eqref{eq:current-collider}.\footnote{For field-theoretic derivations of the conformal collider bounds, see Refs.~\cite{Komargodski:2016gci,Hofman:2016awc,Faulkner:2016mzt}.}

\patchcmd{\thebibliography}{5pt plus 0.2ex minus 0.05ex}{3pt plus 0.2ex minus 0.05ex}{}
  {\PackageError{paper}{Bibliography spacing patch failed}{Check the saved JHEP template.}}
\bibliographystyle{JHEP}
\bibliography{refs}

\end{document}